\documentclass[aps,prd,showpacs,twocolumn,amsmath,10pt,superscriptaddress,floatfix,nofootinbib]{revtex4-1}

\usepackage{epsfig,amssymb,amsfonts,amsmath,mathtools,bm,color,xcolor,graphicx}

\newcommand{\be}{\begin{equation}}
\newcommand{\ee}{\end{equation}}
\newcommand{\bea}{\begin{eqnarray}}
\newcommand{\eea}{\end{eqnarray}}
\newcommand{\ds}{\displaystyle}
\newcommand{\vep}{{\bm p}}
\newcommand{\veq}{{\bm q}}
\newcommand{\Br}{\mbox{Br}}
\newcommand{\BF}{\mbox{BF}}

\newcommand{\gev}{\mbox{GeV}}

\newcommand{\zb}{Z_b^{\pm}(10610)}
\newcommand{\zbp}{Z_b^{\pm}(10650)}
\newcommand{\zc}{Z_{c}^{\pm}(3900)}
\newcommand{\zcp}{Z_{c}^{\pm}(4020)}
\def\vec#1{\boldsymbol{#1}}
\newcommand{\Ne}{N_\text{e}}
\newcommand{\Nin}{N_\text{in}}

\newcommand{\Hb}{\bar{H}}

\newcommand{\nn}{\nonumber}

\graphicspath{{Figures/}}

\begin{document}

\title{The line shapes of the $Z_b(10610)$ and $Z_b(10650)$ in the elastic and inelastic channels revisited}

\author{Q. Wang}
\affiliation{Helmholtz-Institut f\"ur Strahlen- und Kernphysik and Bethe Center for Theoretical Physics, Universit\"at Bonn, D-53115 Bonn, Germany}

\author{V. Baru}
\affiliation{Helmholtz-Institut f\"ur Strahlen- und Kernphysik and Bethe Center for Theoretical Physics, Universit\"at Bonn, D-53115 Bonn, Germany}
\affiliation{Institute for Theoretical and Experimental Physics, B. Cheremushkinskaya 25, 117218 Moscow, Russia}
\affiliation{P.N. Lebedev Physical Institute of the Russian Academy of Sciences, 119991, Leninskiy Prospect 53, Moscow, Russia}

\author{A. A. Filin}
\affiliation{Institut f\"ur Theoretische Physik II, Ruhr-Universit\"at Bochum, D-44780 Bochum, Germany}

\author{C. Hanhart}
\affiliation{Forschungszentrum J\"ulich, Institute for Advanced Simulation, Institut f\"ur Kernphysik and
J\"ulich Center for Hadron Physics, D-52425 J\"ulich, Germany}

\author{A. V. Nefediev}
\affiliation{P.N. Lebedev Physical Institute of the Russian Academy of Sciences, 119991, Leninskiy Prospect 53, Moscow, Russia}
\affiliation{National Research Nuclear University MEPhI, 115409, Kashirskoe highway 31, Moscow, Russia}
\affiliation{Moscow Institute of Physics and Technology, 141700, Institutsky lane 9, Dolgoprudny, Moscow Region, Russia}

\author{J.-L. Wynen}
\affiliation{Forschungszentrum J\"ulich, Institute for Advanced Simulation, Institut f\"ur Kernphysik and
J\"ulich Center for Hadron Physics, D-52425 J\"ulich, Germany}

\begin{abstract}
The most recent experimental data for all measured production and decay channels of the bottomonium-like states $Z_b(10610)$ and 
$Z_b(10650)$ are analysed simultaneously using solutions of the Lippmann-Schwinger equations which respect constraints from
unitarity and analyticity. The interaction potential in the open-bottom channels $B^{(*)}\bar{B}^{*}+\mbox{c.c.}$ contains
short-range interactions as well as one-pion exchange. 
It is found that the long-range interaction does not affect the line shapes as long as only $S$ waves are 
considered. Meanwhile, the line shapes can be visibly modified once $D$ waves, mediated by the strong tensor forces from the pion
exchange potentials, are included. However, in the fit they get balanced largely by a momentum dependent contact term that appears
to be needed also to render the results for the line shapes independent of the cut-off. The resulting line shapes are found to be insensitive
to various higher-order interactions included to verify stability of the results.
 Both $Z_b$ states are found to be described by the poles located on the unphysical Riemann sheets in the vicinity of 
the corresponding thresholds. In particular,
the $Z_b(10610)$ state is associated with a virtual state residing just below the $B\bar{B}^{*}/\bar B{B}^{*}$ threshold while the
$Z_b(10650)$ state most likely is a shallow state located just above the $B^*\bar{B}^{*}$ threshold. 
\end{abstract}



\maketitle

\section{Introduction}

In the last decade, numerous new hadrons have been observed in the charmonium and bottomonium energy region. 
Of special interest are those that cannot be accommodated by the simple quark-model picture and which are, therefore, referred to as exotic hadrons. For example, 
the charged states $\zb$, $\zbp$ \cite{Belle:2011aa}, $\zc$ 
\cite{Ablikim:2013mio,Liu:2013dau}, $\zcp$ \cite{Ablikim:2013wzq}, $Z^\pm(4430)$~\cite{Choi:2007wga,Mizuk:2009da,Chilikin:2013tch,Aaij:2014jqa}
which, amongst other channels, decay into quarkonium states and a pion,
cannot be conventional $\bar{Q}Q$ (with $Q$ denoting the heavy quark) mesons as their minimal quark contents is four-quark. 

The $\zb$ and $\zbp$ bottomonium-like states were observed by the Belle Collaboration as peaks
in the invariant mass distributions of the $\Upsilon(nS)\pi^\pm$ ($n=1,2,3$) and $h_b(mP)\pi^\pm$ ($m=1,2$) subsystems in the dipion transitions from 
the vector bottomonium 
$\Upsilon(10860)$ \cite{Belle:2011aa}. Later, they were confirmed in the elastic 
$B\bar{B}^{*}$\footnote{The quantum numbers of the $Z_b$'s are $J^{PC}=1^{+-}$ \protect{\cite{Collaboration:2011gja}}. Throughout
this paper, a properly normalised $C$-odd combination of the $B\bar{B}^*$ and $\bar{B}B^*$ components is understood.} 
and $B^{*}\bar{B}^{*}$ 
channels \cite{Adachi:2012cx,Garmash:2015rfd}. 
At present, both a tetraquark structure \cite{Ali:2011ug,Esposito:2014rxa,Maiani:2017kyi} and a hadronic molecule interpretation 
\cite{Bondar:2011ev,Cleven:2011gp,Nieves:2011vw,Zhang:2011jja,Yang:2011rp,Sun:2011uh,Ohkoda:2011vj,Li:2012wf,Ke:2012gm,Dias:2014pva}
are claimed to be consistent with the data for these two exotic states.
Their proximity to the $B\bar{B}^*$ and $B^*\bar{B}^*$ thresholds together with the fact that
those are by far the most dominant decay channels 
of the $\zb$ and $\zbp$, respectively, provides a strong support for their
molecular interpretation.
While some works try to explain the $Z_b$'s as a simple kinematical cusps~\cite{Swanson:2014tra}, it was demonstrated in Ref.~\cite{Guo:2014iya} that the 
narrow structures in the elastic channels necessitate near-threshold poles. The general argument presented there
was supported by the explicit analyses presented in Refs.~\cite{Hanhart:2015cua,Guo:2016bjq}. Also the analysis
presented in this work finds poles near the thresholds in conflict with the claims of Ref.~\cite{Swanson:2014tra}.

The literature on the hadronic molecule scenario for the $Z_b$ states is already very rich:
hadronic and radiative decays are studied in 
Refs.~\cite{Voloshin:2011qa,Ohkoda:2012rj,Li:2012uc,Cleven:2013sq,Dong:2012hc,Li:2012as,Ohkoda:2013cea},
the contribution of the two $Z_b$ states to other processes are considered in Refs.~\cite{Chen:2011zv,Chen:2015jgl,Chen:2016mjn}, 
the heavy-quark spin partners of the $Z_b$'s are discussed in 
Refs.~\cite{Mehen:2011yh,Ohkoda:2011vj,Valderrama:2012jv,HidalgoDuque:2012pq,Nieves:2012tt,Guo:2013sya,Karliner:2015ina,Baru:2017gwo},
the line shapes and poles position are addressed in Refs.~\cite{Cleven:2011gp,Hanhart:2015cua,Guo:2016bjq,Kang:2016ezb,Nefediev:2017rfw}, 
predictions of the QCD sum rules are presented in Refs.~\cite{Wang:2013daa,Wang:2014gwa,Agaev:2017lmc} 
and the problem of three-body universality in the $B$-meson sector is investigated in Ref.~\cite{Wilbring:2017fwy}.
For a recent review
of the theory of hadronic molecules we refer to Ref.~\cite{Guo:2017jvc}.
Since both the $\zb$ and $\zbp$ contain a heavy $b\bar{b}$ pair, it is commonly accepted that the 
constraints from the heavy-quark spin symmetry (HQSS) should be very accurate for these systems.
For instance, HQSS allows one to explain naturally the interference pattern in the inelastic channels $\Upsilon(nS)\pi$ and $h_b(mP)\pi$ 
\cite{Bondar:2011ev}. On the contrary, there is a long-lasting debate in the literature about the role of the one-pion exchange (OPE) 
interaction played in the formation of hadronic molecules.
In a pioneering work~\cite{Voloshin:1976ap} a vector meson exchange was proposed
as a key ingredient of the potential between a heavy meson and a heavy antimeson.
In contrast to this the OPE was used in analogy to the deuteron in Refs.~\cite{Tornqvist:1991ks,Tornqvist:1993ng} to predict the
existence of the $X(3872)$. The model was further elaborated and extended to other channels in Refs.~\cite{Liu:2008fh,Thomas:2008ja}, but it was
criticised in Refs.~\cite{Suzuki:2005ha,Kalashnikova:2012qf,Filin:2010se},
where amongst other issues the potential importance of the three-body dynamics was stressed.

A more advanced approach for the $X(3872)$ and other molecular candidates is based on an effective field theory (EFT) treatment which incorporates both short-range interactions parameterised by
a contact term and long-range interactions due to the OPE.\footnote{It is important to notice that in general the OPE potential is field theoretically well defined 
only in connection with a contact operator \cite{Baru:2015nea}.} There exist two competing EFT approaches: one that treats the pions 
as a perturbation on top of the nonperturbative contact interaction --- the so-called X Effective Field Theory 
(X-EFT), see, for example, Refs.~\cite{Voloshin:2003nt,Fleming:2007rp}, and the other in which both the contact and the OPE interactions appear
at leading order in the potential which is
 iterated to all orders fully nonperturbatively 
\cite{Baru:2011rs}. Both approaches can be generalised naturally to the $b$-quark sector. In particular, based on the relatively weak coupling of the pion to heavy fields
in Refs.~\cite{Fleming:2007rp,Valderrama:2012jv} a power counting scheme is proposed within the X-EFT framework to conclude that the central OPE can be included perturbatively in the heavy-quark 
sector.
On the other hand, it was noticed in Refs.~\cite{Baru:2016iwj,Baru:2017gwo} that the most prominent contribution from the OPE 
stems not from the $S$-$S$ but from the $S$-$D$ coupled-channel transitions generated by tensor forces 
and that reliable predictions for the spin partners of the $X$ and $Z_b$'s can only be made if 
the OPE interaction is included nonperturbatively.~\footnote{Note that it is the analogous tensor force
that calls for a non-perturbative treatment of pions in the two-nucleon system~\cite{Fleming:1999ee}.}
For example, due to the OPE both the mass and the width of the $2^{++}$ partner of the $X(3872)$ experience a considerable shift 
as compared to the X-EFT based predictions made in Ref.~\cite{Albaladejo:2015dsa} and the properties of the spin partners of the $Z_b$'s are also quite
sensitive to the pion exchange \cite{Baru:2017gwo} although to a lesser extent than in the $c$-quark sector. 

Within recent phenomenological studies, it was claimed in Ref.~\cite{Voloshin:2015ypa} that, 
near the $S$-wave open-bottom thresholds $B\bar{B}^*$ and $ \bar{B}^*B^*$, the effect of the OPE on the line shapes of the $Z_b$'s can be as large as
 $30\%$. On the other hand, it was advocated in Ref.~\cite{Aceti:2014kja} that the OPE gets cancelled by the one-eta exchange (OEE), so
that in practice it is irrelevant for the formation of the molecular states.
Thus, we take all those controversies as a motivation to investigate in detail the role of the OPE for
the properties of the $Z_b$ states. In particular, in this work we concentrate on the line shapes in the energy region from the $B^*\bar B$ threshold 
to a little above the $B^*\bar B^*$ threshold and extract the poles position of the $Z_b$'s. 
The fact that the full OPE has to be incorporated into a coupled-channel approach to the $B^{(*)}\bar B^{(*)}$ system 
is natural already from the momentum scales involved: 
For energies near the $B^{*}\bar B^{*}$ threshold, the on-shell relative momentum
in the $B^*\bar B$ channel is as large as
\begin{equation}
p_{\rm typ}=\sqrt{m_B\, \delta} \simeq 500~\mbox{MeV},
\label{ptyp}
\end{equation}
where $\delta=m_{B^*}-m_B=45$ MeV denotes the $B^*$-$B$ mass difference, with
$m_B$ and $m_{B^*}$ being the $B$ and $B^*$ meson mass, respectively.
While the $D$ waves in the OPE are indeed suppressed for momenta much smaller than the pion mass ($p\ll m_\pi$), 
in the opposite regime of relevance here ($p_{\rm typ}\gtrsim m_\pi$), 
 $S$-$D$ transitions mediated by the OPE turn out to be as important as $S$-$S$ transitions. 
 
 In this work we analyse the line shapes at
leading order within an EFT approach, which is formulated based on an
 effective Lagrangian consistent with both chiral and heavy-quark
spin symmetries of QCD. The longest range
contribution in the chiral EFT emerges from the
exchanges of the lightest member of the Goldstone-boson octet, the pion. 
Since all the other interactions
are of a shorter range they are integrated out to the order we are working and
included in the EFT Lagrangian via a series of contact interactions with
the low-energy constants (LECs) adjusted to the data. This procedure is in full analogy to the
treatment for few-nucleon systems, see e.g. Ref.~\cite{Epelbaum:2008ga} for a review. 
Thus, in this EFT we treat all the scales such as binding momenta, the pion mass as well as
 the momentum scale $p_{\rm typ}$ as soft while the hard scale 
 $\Lambda_h$ represents a typical chiral EFT breaking scale of the order of $1$ GeV. 
 The scattering (and production) amplitudes are obtained from the nonperturbative solutions of the Lippmann-Schwinger equations (LSEs)
 with the potential which at leading order consists of two momentum independent contact terms and OPE contributions. 
Note that the appearance of a scale as large as $p_{\rm typ}$ in the heavy meson EFT implies that the convergence of the proposed approach, 
which is controlled by the expansion parameter
 \be
 \chi=\frac{p_{\rm typ}}{\Lambda_h}\simeq 0.5,
 \label{chi}
 \ee
 might be relatively slow. Therefore, in what follows we also investigate the impact of
 higher-order interactions such as exchanges of the other members of the SU(3) Goldstone boson octet (that is OEE),
momentum-dependent ${\cal O}(p^2)$ contact terms and HQSS violating 
 contact interactions to understand if their impact on the line shapes is indeed 
 subleading. In the course of this we find that the S-wave-to-D-wave ${\cal O}(p^2)$ contact term is to be promoted to leading order 
 to largely balance the strong modifications of the line shapes generated by the tensor S-wave-to-D-wave contributions from the OPE. 
 It is demonstrated that this promotion is not only necessary to arrive at an acceptable description of the data but
 also to render the results cut off independent.

In Refs.~\cite{Hanhart:2015cua,Guo:2016bjq}, the relevant data were analysed using a 
parameterisation for the line shapes in both elastic and inelastic channels consistent with
unitarity and analyticity based on the leading (${\cal O}(1)$) $S$-wave short-range interactions only. In this paper, we re-analyse 
the same set of data using a direct numerical solution of the LSEs. 
This allows us not only to check the validity 
of the approximations made in Refs.~\cite{Hanhart:2015cua,Guo:2016bjq} in order
to derive self-consistent closed-form analytic expressions but also to
include nonperturbatively the pion exchange and other contributions from the scattering potential.

The paper is organised as follows. In Sect.~\ref{sec:pionless} we discuss the effective potentials in the system at hand. In particular, in Subsect.~\ref{sec:contact}, we start from 
the theory with purely contact transition potentials 
between the elastic and inelastic channels. Then, in Subsect.~\ref{sec:ope}, we discuss the one-pion and one-eta exchange interactions 
between the $B^{(*)}$ mesons and 
include them on top of the contact interactions. The resulting Lippmann-Schwinger equations are derived and solved numerically in 
Sect.~\ref{sec:lse}. Different fitting strategies and the results of the data analysis are presented in Sect.~\ref{sec:fits}. 
Sect.~\ref{sec:poles} is devoted to searches of the poles in the complex plane which describe the $Z_b$ states. 
We summarise in Sect.~\ref{sec:sum}.

\section{Effective potentials}\label{sec:pionless}

The effective potentials between two heavy mesons, which enter LSEs, contain local contact terms and the contributions from the lightest pseudoscalar 
Goldstone boson octet. 
We start from a discussion of the short-range contributions parameterised by the contact potentials without derivatives 
and with two derivatives. Then we discuss in detail the 
one-pion and one-eta exchange potentials relevant for the study. 
In what follows, we stick to the labels introduced previously in Refs.~\cite{Hanhart:2015cua,Guo:2016bjq}: The $\Ne$
open-flavour channels $(\bar{q}b)(\bar{b}q)$ (here $q$ is a light quark) are denoted by
greek letters $\alpha$, $\beta$ and the $\Nin$ hidden-flavour channels $(\bar{b}b)(\bar{q}q)$ --- by latin letters $i$, $j$. 
Elementary poles which would represent compact quark compounds are not considered because the minimal quark contents of the isovector $Z_b$ states is four-quark. In the system at hand, the elastic 
channels are 
$B\bar{B}^*$ and $B^*\bar{B}^*$ and the inelastic channels are $\Upsilon(nS)\pi$ and $h_b(mP)\pi$ with $n=1,2,3$ and $m=1,2$. Therefore, 
$\Ne=2$ and $\Nin=5$.

\subsection{Short-range contributions}\label{sec:contact}

Following the notation introduced above, the full pionless potential takes the form of a $(2\Ne+\Nin)\times(2\Ne+\Nin)$ matrix,
\begin{equation}
V^{\rm pionless}=
\begin{pmatrix}
v_{\alpha\beta}(p,p') & v_{\alpha i}(p,k_i)\\
v_{j\beta}(k'_j,p') & v_{ji}(k'_j,k_i)
\end{pmatrix},
\label{pot}
\end{equation}
where the basis vectors are denoted as 
\be
i=\Upsilon(1S)\pi,~\Upsilon(2S)\pi,~\Upsilon(3S)\pi,~h_b(1P)\pi,~h_b(2P)\pi
\label{inelch}
\ee
and 
\be
\alpha=B\bar{B}^*[S],~ B\bar{B}^*[D],~ B^*\bar{B}^*[S],~ B^*\bar{B}^*[D],
\label{elch}
\ee 
with the letters $S$ or $D$ in square brackets standing for the orbital angular momentum $L$ in the corresponding 
elastic channel. Indeed, the $B\bar{B}^*$ and $B^*\bar{B}^*$ systems with the quantum numbers $J^{PC}=1^{+-}$ can be either in the $^3S_1$ or in the $^3D_1$ state.

In what follows we assume that all inelastic channels only couple to the $S$-wave elastic ones 
as their couplings to the $D$-wave elastic channels are suppressed by a factor $p_\text{typ}^2/m_B^2\ll 1$, 
since the transitions are driven by the exchange of a $B$ meson --- for details we refer to App.~\ref{app:inelD}.
Thus, we set $v_{i\alpha}=0$ for $\alpha = B\bar{B}^*[D]$ and 
$B^*\bar{B}^*[D]$. 
Furthermore, the transition potentials between the $\alpha$-th elastic channel in an $S$ wave and the $i$-th inelastic channel can be parameterised through the 
coupling constants $g_{i\alpha}$,
\begin{eqnarray} 
v_{i\alpha}(k_i,p)=v_{\alpha i}(p,k_i)=g_{i\alpha} k_i^{l_i}, 
\label{via}
\end{eqnarray} 
where $k_i$ and $l_i$ are the momentum and the angular momentum in the $i$-th inelastic channel, respectively. 

The nonvanishing ($S$-wave) couplings $g_{i\alpha}$ are subject to the HQSS 
constraints \cite{Guo:2016bjq,Hanhart:2015cua,Hanhart:2016eyl},
\bea
&\xi_{\Upsilon(nS)}\equiv\frac{g_{[\pi\Upsilon(nS)][B^*\bar{B}^*]}}{g_{[\pi\Upsilon(nS)][B\bar{B}^*]}}=-1,&\nonumber\\[-2mm]
\label{HSconstr}\\[-2mm]
&\xi_{h_b(mP)}\equiv\frac{g_{[\pi h_b(mP)][B^*\bar{B}^*]}}{g_{[\pi h_b(mP)][B\bar{B}^*]}}=1,&\nonumber
\eea
where, as before, $n=1,2,3$ and $m=1,2$. Therefore, in what follows, as long as we discuss the results in the HQSS limit, 
only the coupling constants for the $B\bar{B}^*$ channel are quoted in the form 
\bea
&g_{\Upsilon(nS)} \equiv g_{[\pi\Upsilon(nS)][B\bar{B}^*]},&\nonumber\\[-2mm]
\\[-2mm]
&g_{h_b(mP)} \equiv g_{[\pi h_b(mP)][B\bar{B}^*]}.&\nonumber
\eea
Furthermore, following the arguments from Refs.~\cite{Hanhart:2015cua,Guo:2016bjq}, we neglect the direct interactions in the inelastic channels, since their thresholds are located far away from the 
relevant energy region and the interaction of light states with $\bar QQ$ states is suppressed, 
thus setting $v_{ji}(k',k)=0$ for all $i$'s and $j$'s --- see Eqs.~(\ref{pot}) and (\ref{inelch}) --- that allows us to disentangle the inelastic channels from the elastic ones and to 
reduce the effect of the former to just an additional term in the effective elastic-to-elastic contact transition potential,
\begin{eqnarray} 
V^{\rm CT}_{\alpha\beta}(M,p,p')=v_{\alpha\beta}-\mathcal{G}^{in}_{\alpha\beta}, 
\label{eq:veffective} 
\end{eqnarray}
where 
\begin{eqnarray} 
\mathcal{G}^{in}_{\alpha\beta}=\frac{i}{2\pi M}\sum_i m_{H_i} m_{h_i} g_{i\alpha} g_{i\beta}k_i^{2 l_i+1},
\label{Gindef}
\end{eqnarray} 
with $m_{H_i}$($m_{h_i}$) denoting the mass of the heavy(light) meson in the $i$-th inelastic channel and $M$ being the total energy of the system. Furthermore, the inelastic momentum is defined as 
\begin{eqnarray} 
\label{eq:mominelastic}
k_i&=&\frac{1}{2M}\lambda^{1/2}(M^2,m_{H_i}^2,m_{h_i}^2),\nonumber
\end{eqnarray} 
where $\lambda(m_1^2,m_2^2,m_3^2)$ is the standard triangle function. 

At leading order $O(p^0)$ (LO), the short-range elastic potential $v_{\alpha\beta}$ in the strict HQSS limit consists of two momentum-independent contact interactions 
\cite{AlFiky:2005jd,Bondar:2011ev,Voloshin:2011qa,Mehen:2011yh,Nieves:2012tt} (see also Appendix~\ref{app:Lag} for further details), so that for the basis vectors $\alpha$ (see Eq.~(\ref{elch}))
it can be written as
\begin{eqnarray}
v\left(p,p^\prime \right)=\left(\begin{array}{cccc}
\mathcal{C}_{d}& 0& \mathcal{C}_{f} & 0\\
0 & 0 &0 & 0\\
\mathcal{C}_{f}&0& \mathcal{C}_{d}& 0\\
0 & 0 &0& 0
\end{array}\right).
\label{vSS}
\end{eqnarray}

\begin{widetext}
In what follows, we will also investigate the influence on the line shapes of HQSS violating corrections and next-to-leading order (NLO)
contact interactions with two derivatives. 
To this end, we extend the elastic potential by including three additional contact terms proportional to $\mathcal{D}_d$, $\mathcal{D}_f$ and $\mathcal{D}_{SD} $
at the order ${\cal O}(p^2)$ (see Appendix~\ref{app:Lag} for the corresponding Lagrangian densities), where the first two low-energy constants (LECs) give rise to the $S$-$S$ transitions while 
$\mathcal{D}_{SD}$ contributes to the $S$-$D$ (and $D$-$S$) ones. In addition, we introduce the leading HQSS violating contact interaction $\epsilon$ which contributes to the $S$-$S$ diagonal 
transitions, so 
that for the resulting elastic potential we arrive at
\begin{eqnarray}
\hspace{-0.5cm}v\left(p,p^\prime \right)=\left(\begin{array}{cccc}
\mathcal{C}_{d}(1+\epsilon)+\mathcal{D}_d(p^2+p^{\prime 2}) \hspace{0.3cm} & \mathcal{D}_{SD}p^{\prime2} \hspace{0.3cm} & \mathcal{C}_{f}+\mathcal{D}_f(p^2+p^{\prime 2}) \hspace{0.3cm} & 
\mathcal{D}_{SD}p^{\prime2}\\
\mathcal{D}_{SD}p^{2} & 0 & \mathcal{D}_{SD}p^{2} & 0\\
\mathcal{C}_{f}+\mathcal{D}_f(p^2+p^{\prime 2}) \hspace{0.3cm} & \mathcal{D}_{SD}p^{\prime2} \hspace{0.3cm} & \mathcal{C}_{d}(1-\epsilon)+\mathcal{D}_d(p^2+p^{\prime 2}) \hspace{0.2cm} & 
\mathcal{D}_{SD}p^{\prime2}\\
\mathcal{D}_{SD}p^{2} & 0 & \mathcal{D}_{SD}p^{2} & 0
\end{array}\right).
\label{vfull}
\end{eqnarray}
\end{widetext}

\subsection{One-pion and one-eta exchange potentials}\label{sec:ope}

The potentials for the OPE and the OEE at LO can be derived from the effective
Lagrangian,
\be
\mathcal{L}_{\Phi}=\frac{g_{\scriptscriptstyle Q}}4\mbox{Tr}\left(\vec{\sigma}\cdot\vec{u}_{ab}\bar{H}_b\bar{H}_a^{\dagger}\right)+\mbox{h.c.},
\label{LPhi1}
\ee
where the superfield $H_a$,
\begin{equation}
H_a=P_a+V_a^i\sigma^i,
\end{equation}
describes the heavy-light $\bar{q}Q$ mesons (its transformation properties are discussed in Appendix~\ref{app:Lag}), the axial current is $\vec{u}=-\vec{\nabla}\Phi/f_\pi+\mathcal{O}(\Phi^3)$, 
the matrix for the pseudoscalars reads {(here only the matrix responsible for the SU(2) subspace relevant here is retained for simplicity)}
\be
\Phi=\left(\begin{array}{cc}
\pi^0+\sqrt{\frac 13}\eta & \sqrt2\pi^{+}\\
\sqrt2\pi^{-} & -\pi^0+\sqrt{\frac 13}\eta 
\end{array}\right)
=\pi^i\sigma^i+\sqrt{\frac13}\eta,
\label{pieta}
\ee
and the pion decay constant is $f_\pi=92.4$ MeV \cite{Patrignani:2016xqp}. 

Then, the Lagrangian of the $\Phi$-field coupling to a pair of heavy-light mesons reads, to leading order,
\bea
\mathcal{L}_{\Phi}^{(1)}&=&\frac{g_{\scriptscriptstyle Q}}{2f_\pi}(i\epsilon^{ijk}V_a^{j\dagger}V_b^{k} + V_a^{i\dagger}P_b + P_a^{\dagger}V_b^{i}) \partial^{i}\Phi_{ba}\nonumber\\[-2mm]
\label{eq:Lpi}\\[-2mm]
&+&\frac{g_{\scriptscriptstyle Q}}{2f_\pi}(i\epsilon^{ijk}\bar{V}_a^{j\dagger}\bar{V}_b^{k} 
+\bar{P}_a^{\dagger}V_b^{i}+\bar{V}_a^{\dagger i}\bar{P}_b) 
\partial^{i}\Phi_{ab}.\nonumber
\eea

In the strict heavy-quark limit, the dimensionless coupling $g_{\scriptscriptstyle Q}$ does not depend on the heavy quark flavour $Q$, so that $g_b=g_c$. The latter can be
extracted from the $D^*$ partial decay width,
\begin{eqnarray} 
\Gamma(D^{*+}\to D^0\pi^+)=\frac{g_c^2 m_D k^3}{12\pi f^2_\pi m_{D^*}},
\end{eqnarray} 
where $k$ is the three-momentum in the final state measured in the rest frame of the decaying particle and $m_D$ and $m_{D^*}$ are the $D$- and $D^*$-meson mass, respectively. 
Then, from the width $\Gamma(D^{*+}\to D^0\pi^{+})=56.46~\mathrm{keV}$ \cite{Patrignani:2016xqp} one extracts
\be
g_b=g_c\approx 0.57,
\label{gb}
\ee
which agrees within 10\% with the recent lattice QCD determination of the $B^*B\pi$ coupling constant \cite{Bernardoni:2014kla} and
which will be used in all calculations below.

In this section only the OPE potential will be discussed in detail since the OEE potential 
can be obtained straightforwardly from the expressions for the OPE
by replacing
\begin{itemize}
\item[(i)] the flavour factor +1 (combining the isospin coefficient $-1$ for the OPE in the isovector channel and the negative C-parity factor $-1$) by $-1$ for the OEE;
\item[(ii)] the pion mass, $m_\pi$, by the $\eta$ mass, $m_\eta$;
\item[(iii)] the pion coupling constant $g_b$ by the $\eta$ coupling constant $g_b/\sqrt3$ --- see, for example, 
Ref.~\cite{Grinstein:1992qt} 
and Eq.~(\ref{pieta}) above. Further details can be found in Appendix~\ref{app:eta}.
\end{itemize} 

\begin{figure}[ht] 
\centerline{\epsfig{file=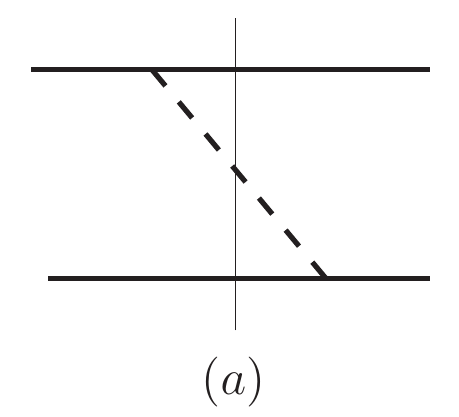, width=0.22\textwidth}\epsfig{file=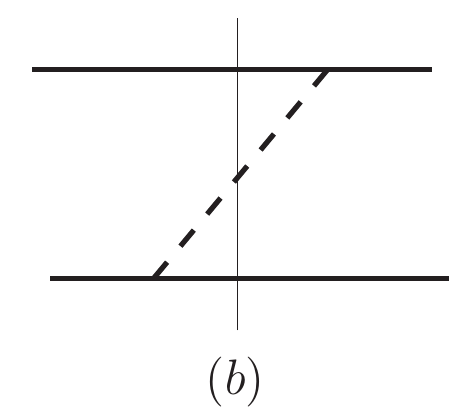, width=0.23\textwidth}}
\caption{Diagrams in the Time-Ordered Perturbation Theory responsible for the two contributions to the OPE potential. The solid and dashed line represent a heavy-light meson and the pion, 
respectively.} 
\label{fig-po-1}
\end{figure}

\begin{figure}[ht]
\centerline{\epsfig{file=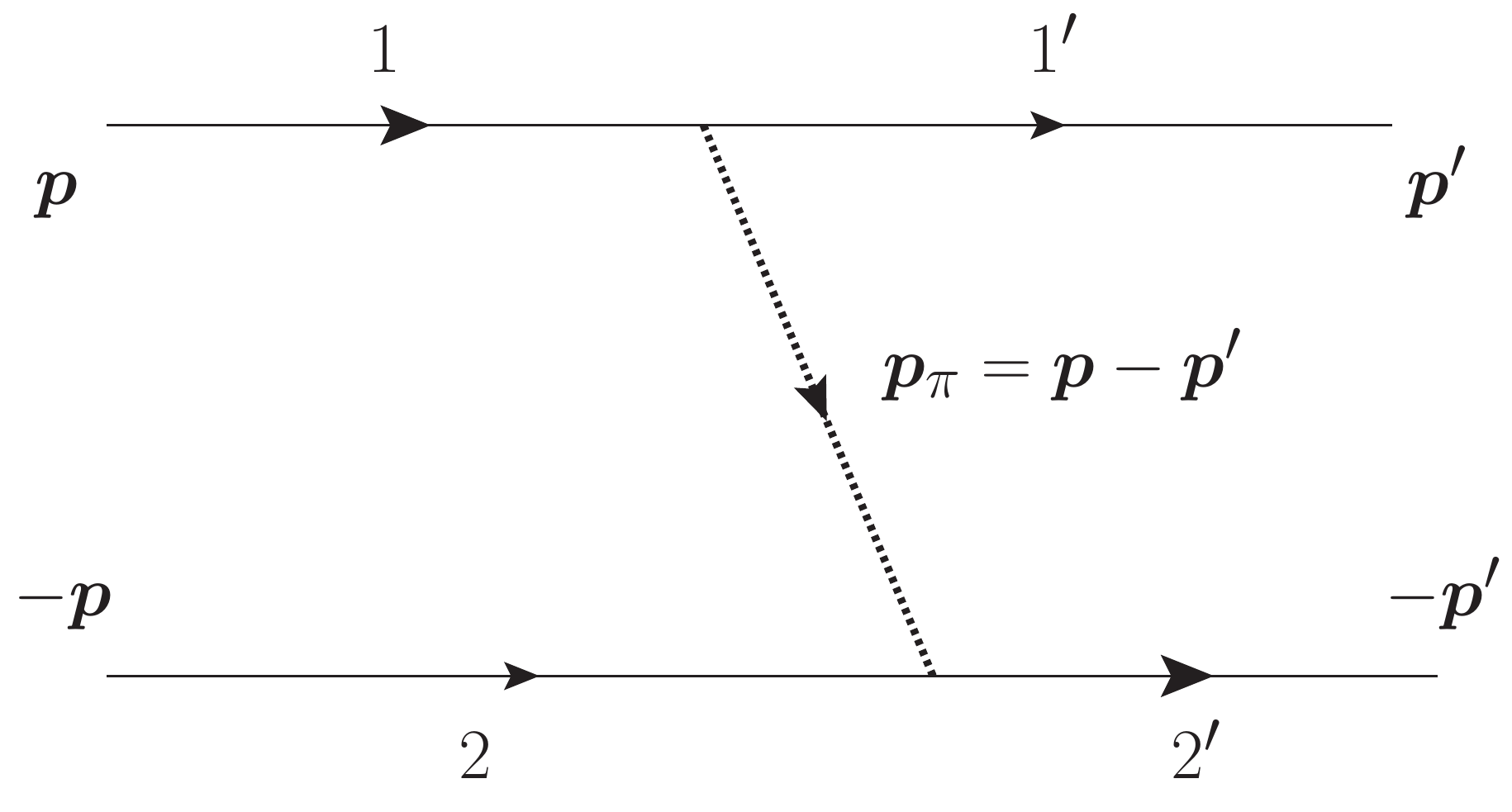, width=0.35\textwidth}} 
\caption{The pion exchange diagram in the $B^{(*)}\bar{B}^*$ channels. The solid and dashed line represent a heavy-light meson and the pion, 
respectively.} 
\label{fig-po-2}
\end{figure}

In the framework of the Time-Ordered Perturbation Theory (TOPT), the OPE potential acquires two contributions depicted in 
Fig.~\ref{fig-po-1}(a) and (b) where the thin vertical line pinpoints the three-body intermediate state. In Fig.~\ref{fig-po-2}, all relevant momenta
are shown explicitly for the first ordering TOPT diagram --- see Fig.~\ref{fig-po-1}(a). Notations for the second diagram depicted in Fig.~\ref{fig-po-1}(b) are
analogous. Then, the OPE potentials at leading order can be written as \cite{Baru:2011rs}
\begin{widetext}
\begin{eqnarray} 
V_a^{nn'}(M,\vep,\vep')&=&-\left(\frac{g_b}{2f_\pi}\right)^2
\frac{p_\pi^np_\pi^{n'}}{2E_\pi(p_\pi)[E_\pi(p_\pi)+E_{1'}(p')+E_2(p)-M]} ,
\label{eq:a}\\ 
V_b^{nn'}(M,\vep,\vep')&=&-\left(\frac{g_b}{2f_\pi}\right)^2\frac{p_\pi^n p_\pi^{n'}}
{2E_\pi(p_\pi)[E_\pi(p_\pi)+E_1(p)+E_{2'}(p')-M]}, 
\label{eq:b} 
\end{eqnarray} 
\end{widetext}
where
\bea
&\ds E_i(p)=m_i+\frac{p^2}{2m_i},\quad i=1^{(\prime)},2^{(\prime)},&\nonumber\\[-2mm]
\\[-2mm]
&E_\pi=\sqrt{\vec p_\pi^2+m_\pi^2}, \quad \vec p_\pi=\vep-\vep',&\nonumber
\eea
with $M$ being the total energy of the system and $E_x$ being the energy of the particle $x$. The incoming and outgoing
momenta are denoted as $\vep$ and $\vep'$, and the Cartesian indices $n$ and $n'$ contract with either the $B^*$ ($\bar{B}^*$) polarisation vectors 
or with the vector product of polarisation vectors, $B^*\times \bar{B}^*$.

In the presence of pions, no additional coupled-channel transitions but those discussed in Sect.~\ref{sec:contact} are possible for 
the $B^{(*)}\bar{B}^*$ systems with the quantum numbers $1^{+-}$. 
Therefore, using the elastic basis vectors introduced in Sect.~\ref{sec:contact} --- see Eq.~(\ref{elch}) --- the OPE potential in this channel can be written as
\begin{eqnarray} 
V^{\pi}(M,p,p')= \left(
\begin{array}{cccc}
V_{SS}^{\pi 11} &V_{SD}^{\pi 11} & V_{SS}^{\pi 12} &V_{SD}^{\pi 12} \\
V_{DS}^{\pi 11} &V_{DD}^{\pi 11} & V_{DS}^{\pi 12} &V_{DD}^{\pi 12} \\
V_{SS}^{\pi 21} &V_{SD}^{\pi 21} & V_{SS}^{\pi 22} &V_{SD}^{\pi 22} \\
V_{DS}^{\pi 21} &V_{DD}^{\pi 21} & V_{DS}^{\pi 22} &V_{DD}^{\pi 22}
\end{array}
\right),
\label{eq:VOPE}
\end{eqnarray} 
where, in each matrix element $V^{\pi\, \lambda\lambda'}_{LL'}(M,p,p')$, the index
$\lambda(\lambda')=1,2$ labels the particle channel ($B\bar{B}^* =1, B^*\bar{B}^*=2$) and $L (L')$ stands for the angular momentum of the state 
$\lambda(\lambda')$. 
The details of the partial 
wave decomposition of the OPE can be found in Ref.~\cite{Baru:2017pvh}; see also Appendix~B of Ref.~\cite{Nieves:2012tt}.
Then,
\begin{widetext}
\begin{eqnarray}\label{eq:OPES}
V_{SS}^{\pi\, \lambda\lambda'}(M,p,p^{\prime})&=&\frac{g_b^2}{24 f_{\pi}^2}\left(\begin{array}{cc}
1 & -2\\
-2& 1
\end{array}
\right)^{\lambda\lambda'}
\Bigl[2pp^{\prime}\Delta_1^{\pi\, \lambda\lambda'}\left(M,p,p^\prime\right)-\left(p^2+p^{\prime2}\right)\Delta_0^{\pi\, \lambda\lambda'}\left(M,p,p^\prime\right)\Bigr],
\nonumber\\
V_{SD}^{\pi\, \lambda\lambda'}(M,p,p^{\prime})&=&\frac{g_b^2}{24\sqrt{2} f_{\pi}^2}\left(\begin{array}{cc}
1 & 1\\
1& 1
\end{array}\right)^{\lambda\lambda'}
\Bigl[3 p^2\Delta_2^{\pi\, \lambda\lambda'}\left(M,p,p^\prime\right)
-4pp^{\prime}\Delta_1^{\pi\, \lambda\lambda'}\left(M,p,p^\prime\right)+\left( 
2p^{\prime2}-p^2\right)\Delta_0^{\pi\, \lambda\lambda'}\left(M,p,p^\prime\right)\Bigr],
\nonumber\\
\\
V_{DS}^{\pi\, \lambda\lambda'}(M,p,p^{\prime})&=&V_{SD}^{\pi\, \lambda\lambda'}(M,p',p),\nonumber\\[2mm]
V_{DD}^{\pi\, \lambda\lambda'}(M,p,p^{\prime})&=&\frac{g_b^2}{24 f_{\pi}^2}
\left[
\left(\begin{array}{cc}
1&0 \\
0&1
\end{array}\right)^{\lambda\lambda'}
V_1^{\pi\, \lambda\lambda'}\left(M,p,p^\prime\right)
-
\left(\begin{array}{cc}
0& 1\\
1&0
\end{array}\right)^{\lambda\lambda'}
V_2^{\pi\, \lambda\lambda'}\left(M,p,p^\prime\right)
\right],
\nonumber
\end{eqnarray}
with 
\begin{eqnarray}
V_1^{\pi\, \lambda\lambda'}\left(M,p,p^\prime\right)&=&\frac92pp^{\prime}\Delta_3^{\pi\, \lambda\lambda'}\left(M,p,p^\prime\right)-3\left(p^2+p^{\prime2}\right)\Delta_2^{\pi\, 
\lambda\lambda'}\left(M,p,p^\prime\right)\nonumber\\
&-&\frac12pp^{\prime}\Delta_1
^{\pi\, \lambda\lambda'}\left(M,p,p^\prime\right)+\left(p^2+p^{\prime2}\right)\Delta_0^{\pi\, \lambda\lambda'}\left(M,p,p^\prime\right),\\
V_2^{\pi\, \lambda\lambda'}\left(M,p,p^\prime\right)&=&\frac92pp^{\prime}\Delta_3^{\pi\, \lambda\lambda'}\left(M,p,p^\prime\right)-\frac32\left(p^2+p^{\prime2}\right)\Delta_2^{\pi\, 
\lambda\lambda'}\left(M,p,p^\prime\right)\nonumber\\
&-&\frac52pp^{\prime}
\Delta_1^{\pi\, \lambda\lambda'}\left(M,p,p^\prime\right)+\frac12\left(p^2+p^{\prime2}\right)\Delta_0^{\pi\, \lambda\lambda'}\left(M,p,p^\prime\right).
\end{eqnarray}
The functions $\Delta_{k}^{\pi\, \lambda\lambda'}$ ($k=0,1,2,3$) are defined as
\be
\Delta_{k}^{\pi\, \lambda\lambda'}(M,p,p') = \int_{-1}^1dx \frac{x^k}{2E_\pi(p_\pi)}\Bigl[D_a^{\pi\, \lambda\lambda'}(p,p',x)+D_b^{\pi\, \lambda\lambda'}(p,p',x)\Bigr],
\label{eq:delta}
\ee
with the contributions of the two TOPT orderings --- see 
Fig.~\ref{fig-po-1} --- given by
\begin{eqnarray}
D_a^{\pi\, \lambda\lambda'}(p,p',x)&=&\frac1{E_\pi(p_\pi)+(E_{1'}(p')+E_2(p))^{\lambda\lambda'}-M}\nonumber\\
&=&\frac1{E_\pi(p_\pi)+(m_{1'}+m_2+p^{\prime 2}/(2m_{1'})+p^2/(2m_2))^{\lambda\lambda'}-M},\label{eq:Da}\\
D_b^{\pi\, \lambda\lambda'}(p,p',x)&=&\frac1{E_\pi(p_\pi)+(E_1(p)+E_{2'}(p'))^{\lambda\lambda'}-M}\nonumber\\
&=&\frac1{E_\pi(p_\pi)+(m_1+m_{2'}+p^2/(2m_1)+p^{\prime2}/(2m_{2'}))^{\lambda\lambda'}-M},\label{eq:Db} 
\end{eqnarray}
where $p_\pi=\sqrt{p^2+p'^2-2pp'x}$, and the mass matrices for the intermediate particles read
\bea
&m_1 =m_{2'} =\left(\begin{array}{cc}
m_{B^*}&m_{B^*} \\
m_{B^*}&m_{B^*}
\end{array}\right),& \nonumber\\
&m_2 =\left(\begin{array}{cc}
m_{B}&m_{B} \\[-2mm]
\\[-2mm]
m_{B^*}&m_{B^*}
\end{array}\right), \quad
m_{1'} =\left(\begin{array}{cc}
m_{B}&m_{B^*} \\
m_{B}&m_{B^*}
\end{array}\right).&\nonumber
\eea 
\end{widetext}
Note that, in the static limit, that is, when the recoil corrections are neglected, the functions $V_1^{\pi\, \lambda\lambda'}$, $V_2^{\pi\, \lambda\lambda'}$, and $D_a^{\pi\, 
\lambda\lambda'}(p,p',x)$, $D_b^{\pi\, 
\lambda\lambda'}(p,p',x)$ 
do not depend on the particle-channel indices ($\lambda$ and $\lambda'$), so that in this approximation 
the dependence on the latter comes entirely from the coefficients in Eq.~(\ref{eq:OPES}).

\section{Numerical solutions of the coupled-channel Lippmann-Schwinger Equations}\label{sec:lse}
 
With the OPE and OEE interaction included and decomposed in partial waves, the full interaction potential in the elastic channels reads
\bea
V^{\rm eff}_{\alpha\beta}(M,p,p')= V^{\rm CT}_{\alpha\beta}(M,p,p') &+ &V^{\pi}_{\alpha\beta}(M,p,p')\nonumber\\[-2mm]
\\[-2mm]
&+&V^{\eta}_{\alpha\beta}(M,p,p'),\nonumber
\label{pot2}
\eea
where the effects of the inelastic channels are contained in $V^{\rm CT}_{\alpha\beta}(M,p,p') $ as given by Eq.~\eqref{eq:veffective}
and the channel indices are defined by Eqs.~(\ref{inelch}) and (\ref{elch}).
 
We work in terms of the production amplitudes 
rather than the scattering amplitudes which are more convenient given that we aim at fits for the invariant mass distributions
measured in the $\Upsilon(10860)$ decays. Thus, the system of the LSEs reads
\bea
U_\alpha(M,p)&=&F_\alpha(M,p)\nonumber\\[-1mm]
\label{eq:LSE}\\[-1mm]
&-&\sum_\beta\int
U_\beta(M,q)G_\beta(M,q)V_{\beta\alpha}^{\text{eff}}(M,q,p)\frac{d^3q}{(2\pi)^3}.\nonumber
\eea
Here $F_\alpha(M,p)$ and $U_\alpha(M,p)$ denote the Born and the physical production amplitude of the $\alpha$-th elastic channel
from a point-like source, respectively.
We include those source terms in $S$ waves only. Note that a $D$ wave contribution to the source could be re-expressed
as an additional energy dependence of the source, which, however, is not required by the data.
This is
also in line with the results of Ref.~\cite{Mehen:2013mva} from which one concludes that the two-step process $\Upsilon(10860)\to B^*B^{(*)}\to B^*B^{(*)}\pi$
with the pion emitted from the $B$-meson line, that provides the energy dependence, is suppressed as compared to 
a point-like source.
The Green's function for a two-heavy-meson intermediate state reads
\be
G_\alpha(M,q)=\frac{2\mu_\alpha}{q^2-p_\alpha^2-i\epsilon},\quad p_\alpha^2\equiv 2\mu_\alpha(M-m_{\rm th}^\alpha),
\ee 
where $m_{\rm th}^\alpha$ stands for the $\alpha$-th elastic threshold and $\mu_{\alpha}$ is the reduced mass in the channel $\alpha$. Other components of the multichannel amplitude responsible 
for the 
production of the inelastic channels in the final state can be obtained 
from $U_\alpha(M,p)$ algebraically which is a consequence of the omitted direct interactions in the inelastic 
channels~\cite{Hanhart:2015cua,Guo:2016bjq}. In particular, for the $i$-th inelastic channel in the final state we have
\be
U_i(M,p_i)=-\int\frac{d^3q}{(2\pi)^3}U_\alpha(M,q) G_\alpha(M,q)v_{\alpha i}(M,q,p_i),
\label{eq:inelastic}
\ee
$$
p_i=\frac{1}{2M}\lambda^{1/2}(M^2,m_{1i}^2,m_{2i}^2),
$$
where $\lambda(m_1^2,m_2^2,m_3^2)$ is the standard triangle function. 
Expression (\ref{eq:inelastic}) is based on the assumption that the data are dominated by the $Z_b(10610)$ and $Z_b(10650)$ poles which emerge from $B^{(*)}\bar B^*$ dynamics, so that the Born 
amplitudes $F_i(M,p)$ coming from the inelastic sources can be safely neglected. 
While this is well justified for the $h_b(mP) \pi$ channels, since the $Z_b$ poles are necessary for the change in the heavy quark spin,
it appears to be unjustified for the 
heavy-spin-conserving $\Upsilon(nS)\pi$ channels. However, to fully control the interplay of the source term and the 
resonance terms one would need to properly include the $\pi\pi$ interaction 
which goes beyond the scope of this work.\footnote{Relevant calculations for the decays of the $\Upsilon(3S)$ and $\Upsilon(4S)$ 
are presented in Refs.~\cite{Chen:2015jgl,Chen:2016mjn}. However, extending those calculations
to the decay of $\Upsilon(5S)$ is technically very demanding due to a more complicated
analytic structure of the transition matrix elements.}
This is why, in what follows, we do not include the line shapes in the $\Upsilon(nS)\pi$ channels into the fit. 

In terms of the production amplitudes the expressions for
the differential widths in the elastic and inelastic channels read (see Refs.~\cite{Hanhart:2015cua,Guo:2016bjq} for the derivation)
\begin{eqnarray}
\frac{d\Gamma_\alpha}{dM}&\equiv&y_\alpha^{\rm th}(M)=\frac13\ \frac{2m_{B^{(*)}}2m_{B^*}2m_{\Upsilon(10860)}}{32\pi^3 m_{\Upsilon(10860)}^2}\; p_\pi^* p_\alpha |U_\alpha|^2,\nonumber\\[-2mm]
\\[-2mm]
\frac{d\Gamma_i}{dM}&\equiv&y_i^{\rm th}(M)=\frac13\ \frac{2m_{h_i}2m_{H_i}2m_{\Upsilon(10860)}}{32\pi^3 m_{\Upsilon(10860)}^2}\; p_\pi^* p_i|U_i|^2,\nonumber
\end{eqnarray}
respectively, where $p_\pi^*$ is the three-momentum of the spectator pion in the rest frame of the $\Upsilon(10860)$ and $p_{\alpha}(p_i)$ is the 
three-momentum in the $\alpha$-th elastic ($i$-th inelastic) channel in the rest frame of the $B^*\bar{B}^{(*)}$ ($\Upsilon(nS)\pi/h_b(mP)\pi$) system. 
Then, the total branching in an elastic or inelastic channel $x$ is defined as 
\be
\Br_x=\frac{\Gamma_x}{\sum_{\alpha=1}^{\Ne}\Gamma_\alpha+\sum_{i=1}^{\Nin}\Gamma_i},
\label{Brx}
\ee
where
\be
\Gamma_x=\int_{M_{\rm min}}^{M_{\rm max}}y_x^{\rm th}(M)dM.
\ee
 For simplicity, we define the branching fractions (BFs) relative to the $B\bar{B}^*\pi$ channel,
\be
\mbox{BF}_x=\frac{\Br_x}{\Br_{B\bar{B}^*\pi}}=\frac{\Gamma_x}{\Gamma_{B\bar{B}^*\pi}}.
\label{BFs}
\ee
The $\chi^2$ function to be minimised in the fitting process is built as
\begin{widetext}
\bea
\chi^2&=&\sum_{\alpha=B^{(*)}\bar{B}^*}\sum_{n_\alpha}\left(\frac{{\cal N}_\alpha y^{\rm th}_\alpha(M_{n_\alpha})-y^{\rm exp}_\alpha(M_{n_\alpha})}{\delta_{n_\alpha}}\right)^2
+
\left(\frac{\BF_{B^{*}\bar{B}^*}^{\rm th}-\BF_{B^{*}\bar{B}^*}^{\rm exp}}{\delta_{{\rm BF}_{B^{*}\bar{B}^*}}}\right)^2 \nonumber\\
&+&\sum_{i=h_b(mP)\pi}\left\{\sum_{n_i}\left(\frac{{\cal N}_i y^{\rm th}_i(M_{n_i})-y^{\rm exp}_i(M_{n_i})}{\delta_{n_i}}\right)^2+
\left(\frac{\BF_i^{\rm th}-\BF_i^{\rm exp}}{\delta_{{\rm BF}_i}}\right)^2\right\}\label{chi2}\\
&+&\sum_{j=\Upsilon(nS)\pi}
\left(\frac{\BF_j^{\rm th}-\BF_j^{\rm exp}}{\delta_{{\rm BF}_j}}\right)^2,\nonumber
\eea
\end{widetext}
where $y^{\rm exp}$'s are the experimental distributions given in the form of histograms (the sums in $n$'s run over bins), $\delta$'s denote
the errors, and
the normalisation factors ${\cal N}_x$ are additional (auxiliary) fitting parameters since the data are presented in arbitrary units. 
Both line shapes and total 
branchings are used in the fit for the elastic $B^{(*)}\bar{B}^*$ and the inelastic $h_b\pi$ channels while for the inelastic channels $\Upsilon\pi$ 
only the total branchings can be used in the one-dimensional analysis.

Since the production proceeds via pion emission from the $\Upsilon(10860)$ bottomonium, the corresponding Born amplitudes are subject to a HQSS constraint 
similar to the one given in Eq.~(\ref{HSconstr}) for the coupling constants, that is, $\xi_{\Upsilon(10860)}\equiv F_{B^*\bar{B}^*}(M,p)/F_{B\bar{B}^*}(M,p)=-1$ \cite{Hanhart:2015cua,Guo:2016bjq}.
Then, without loss of generality, one can set 
\be
F_{B\bar{B}^*}(M,p)=-F_{B^*\bar{B}^*}(M,p)=1,
\ee
to fix the overall normalisation of the amplitudes which in any case drops out from the branchings and can be absorbed by the unimportant factors ${\cal N}$ in the $\chi^2$ function --- see 
Eqs.~(\ref{Brx}) and (\ref{chi2}). 

\begin{widetext}
We may now perform the angular integrations in Eq.~(\ref{eq:LSE}) to reduce the three-dimensional integral equation to a one-dimensional equation,
\be
U_\alpha(M,p)=F_\alpha(M,p)-\frac1{\pi^2}\sum_\beta\mu_\beta \int_0^\infty \frac{q^2 U_\beta(M,q)V_{\beta\alpha}^{\text{eff}}(M,q,p)}{q^2-p_\beta^2-i\epsilon}dq.
\ee
To render the integrals well defined we introduce a sharp ultraviolet cut-off $\Lambda$
which needs to be larger than all typical three-momenta related to
the coupled-channel dynamics.
Unless stated otherwise, for the results presented below we choose $\Lambda=1$~GeV. The cut-off dependence of our results will be discussed in Sect.~\ref{sec:renorm}.
 When the energy goes above the threshold of the intermediate channel $\beta$,
the on-shell three-momentum $p_\beta$ is real allowing for a singular integrand at $q^2=p_\beta^2$.
In this case one subtraction at the point $q=p_\beta$ is implemented to stabilise the numerical result,
\begin{eqnarray}
U_\alpha(M,p)&=&F_\alpha(M,p)\nonumber\\
&-&\frac1{\pi^2}\sum_\beta\mu_\beta\int_0^\Lambda \frac{q^2U_\beta(M,q) V_{\beta\alpha}^{\text{eff}}(M,q,p)
-p_\beta ^2U_\beta (M,p_\beta )V^{\text{eff}}_{\beta\alpha}(M,p_\beta ,p)}{q^2-p_\beta ^2-i\epsilon}dq\nonumber \\
&-&\frac1{2\pi^2}\sum_\beta\mu_\beta p_\beta U_\beta (M,p_\beta )V_{\beta\alpha}^{\text{eff}}(M,p_\beta ,p)\left(i\pi- 
\log\left(\frac{\Lambda+p_\beta }{\Lambda-p_\beta }\right)\right).
\end{eqnarray}
\end{widetext}
Since this equation is valid in the whole complex energy plane, it is also used to find the resonance poles in Sect.~\ref{sec:poles}.

The masses of particles used in the calculations are \cite{Patrignani:2016xqp}
\bea
&&m_B=5279~{\rm MeV},\nonumber\\
&&m_{B^*}=5324~{\rm MeV},\nonumber\\
&&m_\pi=137.28~{\rm MeV},\nonumber\\ 
&&m_{\Upsilon(10860)}=10860~{\rm MeV},\nonumber\\
&&m_{\Upsilon(1S)}=9460~{\rm MeV},\\
&&m_{\Upsilon(2S)}=10023~{\rm MeV},\nonumber\\
&&m_{\Upsilon(3S)}=10355.2~{\rm MeV},\nonumber\\
&&m_{h_b(1P)}=9898.6~{\rm MeV},\nonumber\\
&&m_{h_b(2P)}=10259.8~{\rm MeV}.\nonumber
\eea

\section{Fit schemes and results}\label{sec:fits} 

In this section we fit the line shapes of the two $Z_b$ states in both elastic ($B^{(*)}\bar{B}^*$) and inelastic ($h_b(mP)\pi$, $m=1,2$) channels. 
As was already mentioned above, the line shapes in the inelastic $\Upsilon(nS)\pi$ ($n=1,2,3$) channels cannot be included into the fit yet, since the data contain a significant contribution 
driven by the two-pion final state interaction that we cannot include straightforwardly in the present approach.
In addition, the analysis has to be multidimensional. 
Meanwhile, the partial branchings for all 
measured elastic and inelastic channels are included in our analysis --- see Eq.~(\ref{chi2}) for the formula for $\chi^2$ and Table~\ref{tab:fractions}
where the branchings are quoted relative to the 
$B\bar{B}^*\pi$ channel. In the branching fractions the interference terms with crossed channels are effectively included
via scaling factors gauged in test calculations. It turns out that those scaling factors are very close to 1 in the
$h_b\pi$ channels, but are $5\%$, $10\%$, and $20\%$ for the partial BF's in the $\Upsilon(1S)\pi$, $\Upsilon(2S)\pi$, and $\Upsilon(3S)\pi$ channels, respectively.

\begin{table*}[t!]
\begin{center}
\begin{tabular}{|c|c|c|c|c|c|c|}
\hline 
BF, \% & $B^{*}\bar{B}^{*}\pi$ & $\Upsilon(1S)\pi\pi$ & $\Upsilon(2S)\pi\pi$ & $\Upsilon(3S)\pi\pi$ & $h_{b}(1P)\pi\pi$ & $h_{b}(2P)\pi\pi$\tabularnewline
\hline 
\hline 
Exp. & $50\pm10$ & $0.6\pm0.3$ & $4\pm1$ & $2\pm1$ & $9\pm2$ & $15\pm3$\tabularnewline
\hline 
A & $58.04_{-5.83}^{+6.00}$ & $0.55_{-0.24}^{+0.34}$ & $3.34_{-1.29}^{+1.76}$ & $1.72_{-0.82}^{+1.17}$ & $8.25_{-2.37}^{+3.22}$ & $11.52_{-3.73}^{+4.73}$\tabularnewline
\hline
G & $54.13_{-18.07}^{+18.83}$ & $0.55_{-0.26}^{+0.41}$ & $3.51_{-1.48}^{+2.28}$ & $1.83_{-0.95}^{+1.59}$ & $9.18_{-2.40}^{+3.59}$ & $14.92_{-4.09}^{+6.00}$\tabularnewline
\hline 
\end{tabular}
\end{center}
\caption{The branching fractions (BFs), in per cent, for the elastic and inelastic channels in the $\Upsilon(10860)$ decays via the two $Z_b$ states relative to the $B\bar{B}^*\pi$ channel for 
which the BF is set to unity~\cite{Garmash:2014dhx,Belle:2011aa,Garmash:2015rfd}.}
\label{tab:fractions} 
\end{table*}

\subsection{The role of pion dynamics, HQSS violation and higher-order interactions}

\begin{figure*}
\begin{center}
\epsfig{file=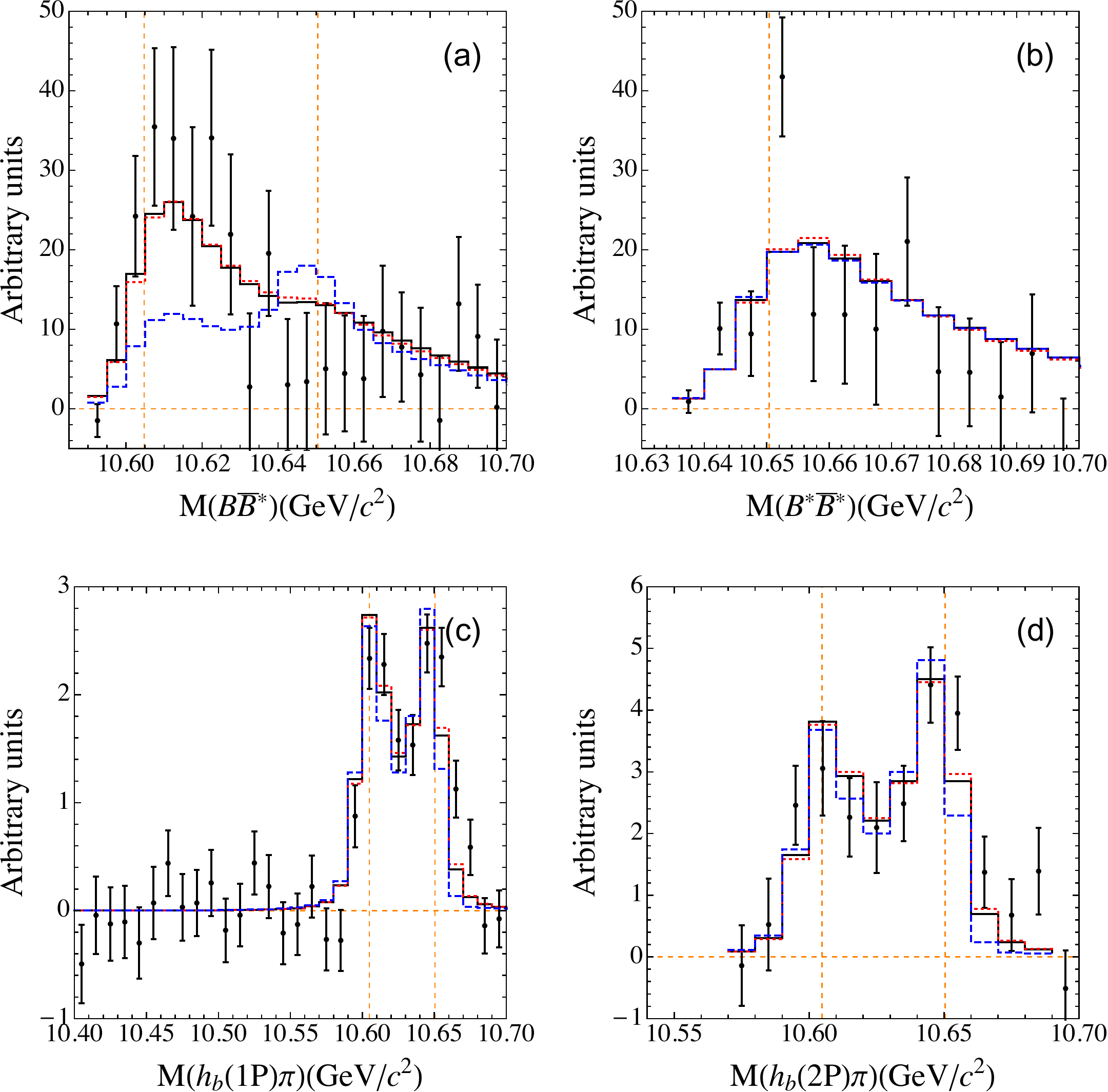, width=0.75\textwidth}
\caption{\label{fig:abc} The fitted line shapes for Schemes A (solid black curves), B (red dotted curves) and C (blue dashed curves)
in the elastic $B^{(*)}\bar{B}^*$ and
in the inelastic $h_b(mP)\pi$ ($m=1,2$) channels. 
The vertical dashed orange lines indicate the position of the $B\bar{B}^*$ and $B^*\bar{B}^*$ thresholds. The experimental data are from Refs.~\cite{
Belle:2011aa,Garmash:2015rfd}.}
\end{center}
\end{figure*}

In this section, we investigate 
the impact of various effects, included in this work for the first time, on the line shapes. To gain a proper insight on the role of these effects, we include them one by one. We, therefore, 
consider the following different schemes:
\begin{itemize}
\item {\bf Scheme A}: Only the $\mathcal{O}(p^0)$ $S$-wave contact potentials are considered. 
\item {\bf Scheme B}: As Scheme A with OPE added, however, only in $S$ waves.
\item {\bf Scheme C}: As Scheme B, but with $D$-wave OPE included.
\item {\bf Scheme D}: As Scheme C but allowing for a sizeable HQSS violation.
\item {\bf Scheme E}: As Scheme C, but with the $\mathcal{O}(p^2)$ $\mathcal{D}_{SD}$ contact potential. 
\item {\bf Scheme F}: As Scheme C, but with the $\mathcal{O}(p^2)$ $\mathcal{D}_d$, $\mathcal{D}_f$, $\mathcal{D}_{SD}$ contact potentials.
\item {\bf Scheme G}: As Scheme F, but with OEE included in addition.
\end{itemize}

\begin{table*} 
\begin{tabular}{|c|c|c|c|c|c|c|}
\hline 
Scheme & $|g_{\Upsilon(1S)}|$ & $|g_{\Upsilon(2S)}|$ & $|g_{\Upsilon(3S)}|$ & $|g_{h_{b}(1P)}|$ & $|g_{h_{b}(2P)}|$ & $\chi^{2}/\text{d.o.f.}$\tabularnewline
\hline 
\hline 
A & $0.30\pm0.07$ & $1.01\pm0.20$ & $1.28\pm0.34$ & $3.29\pm0.38$ & $11.38\pm 1.46$ & $1.29$\tabularnewline
\hline 
B & $0.31\pm0.07$ & $1.05\pm0.20$ & $1.33\pm0.35$ & $3.19\pm0.35$ & $11.15\pm1.40$ & $1.23$\tabularnewline
\hline 
C & $0.30\pm0.07$ & $1.00\pm0.20$ & $1.22\pm0.33$ & $4.85\pm0.64$ & $16.24\pm2.54$ & $2.00$\tabularnewline
\hline 
D & $0.38\pm0.55$ & $1.35\pm1.08$ & $1.89\pm0.52$ & $3.94\pm0.52$ & $13.87\pm1.98$ & $1.54$\tabularnewline
\hline 
E & $0.30\pm0.07$ & $1.04\pm0.20$ & $1.43\pm0.36$ & $2.76\pm0.29$ & $9.99\pm1.08$ & $0.95$\tabularnewline
\hline 
F & $0.26\pm0.06$ & $0.88\pm0.17$ & $1.16\pm0.29$ & $1.89\pm0.22$ & $6.98\pm0.81$ & $0.83$\tabularnewline
\hline 
G & $0.25\pm0.06$ & $0.88\pm0.17$ & $1.15\pm0.29$ & $1.92\pm0.22$ & $7.07\pm0.84$ & $0.83$\tabularnewline
\hline
\end{tabular}
\caption{\label{tab:par-1}The fitted values of the coupling constants for Schemes A-H. The cut-off $\Lambda$ is set to 1~GeV as discussed in the 
text. The couplings $g_{\Upsilon(nS)}$ ($n=1,2,3$)
are given in the units of $\gev^{-2}$ while the couplings $g_{h_b(mP)}$ ($m=1,2$) have the dimension of $\gev^{-3}$. Only the absolute values are presented 
since physical quantities are not sensitive to the couplings' signs. To demonstrate the quality of each fit we quote the corresponding 
reduced $\chi^2/\text{d.o.f.}$ in the last column.} 
\end{table*}

\begin{table*} 
\begin{tabular}{|c|c|c|c|c|}
\hline 
Scheme & $\mathcal{C}_{d}$ & $\mathcal{C}_{f}$ & $\mathcal{D}_{d}$ & $\mathcal{D}_{SD}$\tabularnewline
\hline 
\hline 
A & $-3.30\pm0.11$ & $-0.06\pm0.13$ & $0$ & $0$\tabularnewline
\hline 
B & $-0.51\pm 0.11$ & $-5.64\pm0.13$ & $0$ & $0$\tabularnewline
\hline 
C & $0.80\pm0.14$ & $-4.50\pm0.15$ & $0$ & $0$\tabularnewline
\hline 
D & $1.22\pm0.20$ & $-4.71\pm0.21$ & $0$ & $0$\tabularnewline
\hline 
E & $-0.08\pm0.30$ & $-4.15\pm0.59$ & $0$ & $-5.83\pm0.57$\tabularnewline
\hline 
F & $1.63\pm0.43$ & $-5.28\pm0.26$ & $-2.99\pm0.58$ & $-3.93\pm0.53$ \tabularnewline
\hline 
G & $1.34\pm0.40$ & $-3.95\pm0.27$ & $-3.38\pm0.54$ & $-3.13\pm0.61$\tabularnewline
\hline
\end{tabular}
\caption{
\label{tab:par-2} The $\mathcal{O}(p^0)$ ($\mathcal{C}_{d}$ and $\mathcal{C}_{f}$) and $\mathcal{O}(p^2)$ ($\mathcal{D}_{d}$ and 
$\mathcal{D}_{SD}$)
contact terms 
(in the units of $\mathrm{GeV}^{-2}$ and $\mathrm{GeV}^{-4}$, respectively) for each fit scheme. The contact term $\mathcal{D}_{f}$ is set to 0 because 
it is strongly correlated with $\mathcal{C}_{f}$, does not affect the value of $\chi^2/\text{d.o.f.}$ and is, therefore, redundant.} 
\end{table*}
The parameters of the fits for all schemes described above are listed in Tables~\ref{tab:par-1} and ~\ref{tab:par-2}. 
In what follows, we discuss the quality of the fits and draw conclusions on the role played by various effects. 

The fit results for Schemes A, B and C are shown in Fig.~\ref{fig:abc} by the solid black, red dotted and blue dashed curves, respectively.
The line shapes of Scheme A
are basically identical to those based on the parameterisation proposed in Refs.~\cite{Guo:2016bjq,Hanhart:2015cua,Hanhart:2016eyl}. 
In particular, also here the diagonal and off-diagonal matrix elements of the direct interaction potential satisfy the strong inequality 
$|\mathcal{C}_f|\ll |\mathcal{C}_d|$ which ensures that the transitions between the two elastic channels are suppressed.
It has to be noticed, however, that
the parameters in Scheme A of the present paper cannot be directly compared with those from Ref.~\cite{Guo:2016bjq} because of a different regularisation. 

Scheme B, with the $S$-wave dynamic OPE included, provides a fit of the same quality as the one in Scheme A. The resulting line shapes are quite similar for both fits
which is consistent with the claim made in Ref.~\cite{Guo:2016bjq} on a moderate role played by the OPE. On the contrary, the claim of 
Ref.~\cite{Voloshin:2015ypa} that already the $S$-wave OPE changes the line shape up to $30\%$ near threshold is not supported by our fit B. 
It is instructive to 
identify the most relevant differences between the 
approaches employed in Ref.~\cite{Voloshin:2015ypa} and in this work: First, 
the off-diagonal terms connecting the $B\bar B^*$ with the $B^*\bar B^*$ channel and neglected in Ref.~\cite{Voloshin:2015ypa}
turn out to be large in the OPE potential and even exceed the diagonal terms by roughly a factor 
of 2 --- see Eq.~(\ref{eq:OPES}). Secondly, the argument of Ref.~\cite{Voloshin:2015ypa} is based on a perturbative treatment of the OPE while in this
work the pions are treated nonperturbatively and all parameters are re-fitted to the data. Indeed, before drawing conclusions on the role of long-range 
interactions the short-range part of the OPE, intimately connected to the contact interactions \cite{Baru:2015nea}, 
needs to be appropriately renormalised which is achieved in this work by refitting the contact terms to the line shapes. 
After the re-fit we find that the central $S$-wave OPE can be absorbed to a very large extent into a re-definition of the contact 
interactions which is in line with the observation made in Ref.~\cite{Baru:2017gwo}.

\begin{figure*}
\begin{center}
\epsfig{file=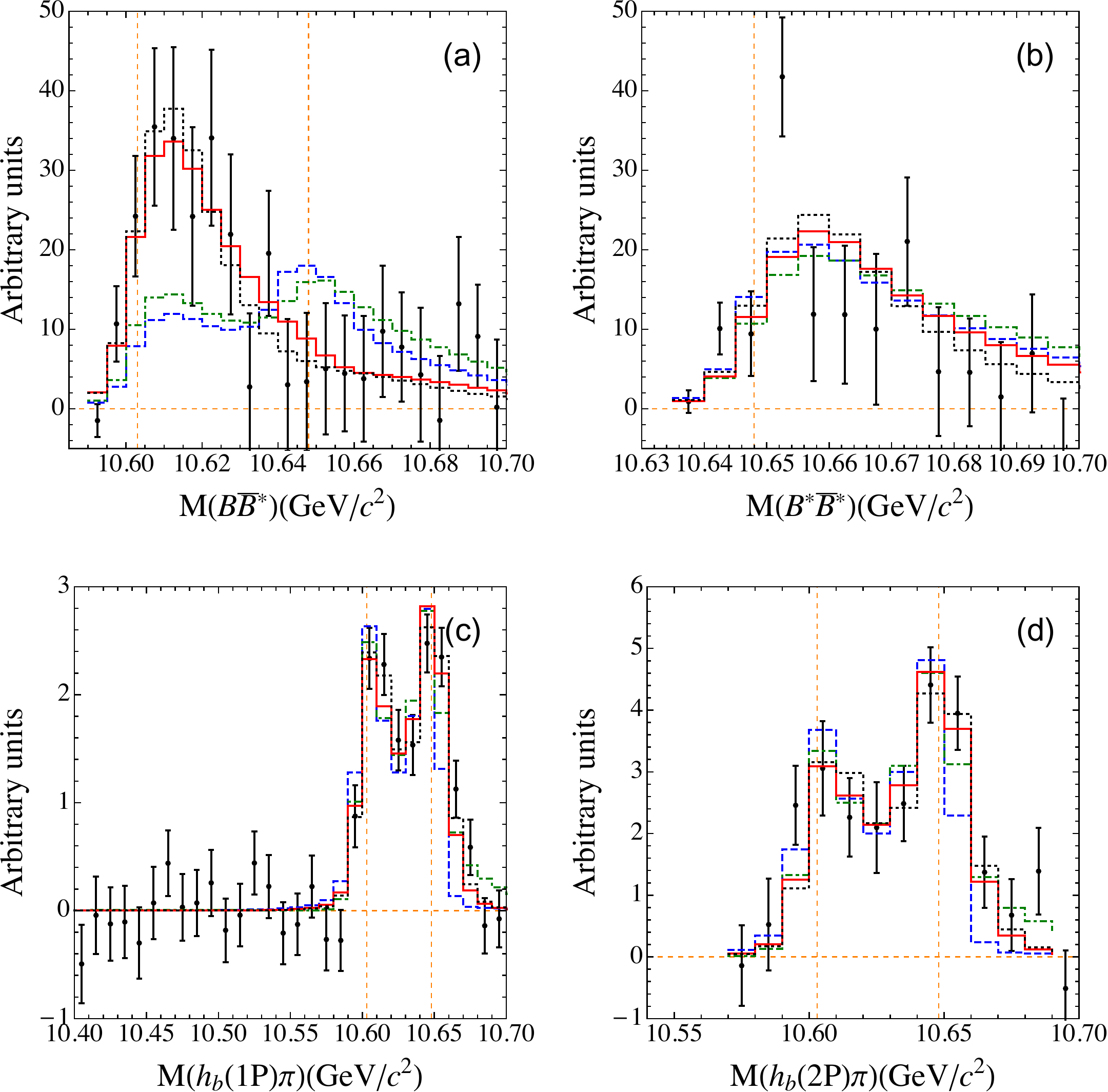, width=0.8\textwidth}
\caption{ \label{fig:cdef} The fitted line shapes for Schemes C (blue dashed curves), D (green dot-dashed curves), E (red solid curves) and G 
(black dotted curves)
in the elastic $B^{(*)}\bar{B}^*$ and
in the inelastic $h_b(mP)\pi$ ($m=1,2$) channels. The experimental data are from Refs.~\cite{
Belle:2011aa,Garmash:2015rfd}.}
\end{center}
\end{figure*} 

The fit for Scheme C demonstrates that $D$ waves in the OPE can play a non-negligible role for the line shapes --- this is most clearly 
visible in the $B\bar B^*$ spectra in Fig.~\ref{fig:abc} where now a bump appears around the $B^*\bar B^*$ threshold. 
This result should not come as a surprise given the large momentum scale $p_{\rm typ}$, defined in Eq.~(\ref{ptyp}), introduced by the 
splitting between the elastic thresholds, which enhances the $D$ waves and, in particular, the contribution from the $S$-$D$ transitions. 
These findings are in line with the claims made in Refs.~\cite{Baru:2016iwj,Baru:2017gwo}.
Although the fit for Scheme C is essentially consistent with the $B^*\bar B^*$ distribution as well as with the distributions in the inelastic channels, 
the shape of the $B\bar B^*$ spectrum distorted by the $D$-wave OPE is not supported by the data. 
The observation that the experimental line shapes do not exhibit a hump structure around the $B^*\bar B^*$ threshold was related in Ref.~\cite{Voloshin:2016cgm}
with a possible existence of the light-quark symmetry in QCD.
In what follows, we investigate two different variations of the potential aiming at an improved description of the data.

In Fig.~\ref{fig:cdef} we demonstrate the impact of various higher-order interactions on the line shapes. 
In particular, in the fit for Scheme D (green dot-dashed curves) we release the HQSS breaking parameters in the 
elastic and inelastic potentials as well as in the production vertex (that is, $\xi_{\Upsilon(nS)}$, $\xi_{h_b(mP)}$ and $\epsilon$) and allow them to deviate up to 50\%
from the HQSS predicted values. This gives
\bea
&&\xi_{\Upsilon(10860)}=-1.23\pm0.07,\nonumber\\
&&\xi_{\Upsilon(1S)}=-0.66\pm0.82,\nonumber\\
&&\xi_{\Upsilon(2S)}=-0.61\pm0.52,\nonumber\\ 
&&\xi_{\Upsilon(3S)}=-0.50\pm0.54,\\ 
&&\xi_{h_b(1P)}=1.50\pm0.07,\nonumber\\ 
&&\xi_{h_b(2P)}=1.50\pm0.63,\nonumber\\
&&\epsilon=-0.50\pm0.04.\nonumber
\eea

However, in spite of a significant HQSS breaking allowed in the fit, the resulting distribution does not show 
a qualitative improvement leaving, in particular, the bump structure around the $B^*\bar B^*$ threshold nearly unchanged. 

On the other hand, the inclusion of a single ${\cal O}(p^2)$ contact interaction $\mathcal{D}_{SD}$ between the
$S$-wave and $D$-wave elastic channels (Scheme E) improves the fit considerably (see the red solid curves in Fig.~\ref{fig:cdef}) 
yielding $\chi^2/\mbox{d.o.f} =0.95$, as given in Table~\ref{tab:par-1}. As required by the data, this term is fine-tuned to cancel a large 
portion of the $S$-$D$ contribution generated by the tensor part of the OPE. 
Further, the inclusion of the $\mathcal{O}(p^2)$ counter terms $\mathcal{D}_{d}$ and
$\mathcal{D}_{f}$ between the $S$-wave elastic channels in addition to the $\mathcal{D}_{SD}$ term results only in a very minor change in the fit
with the $\chi^2/\mbox{d.o.f} =0.83$. In addition, we observe that the low-energy constant $\mathcal{D}_f$ is by almost 100\% correlated with $\mathcal{C}_f$, so that by setting 
$\mathcal{D}_f=0$ one gets the fit with almost exactly the same $\chi^2/\mbox{d.o.f}$. By adding also the OEE interaction (Scheme G) one obtains results which lie
on top of those for Scheme F (see the black dotted curves for Scheme G) 
in Fig.~\ref{fig:cdef}.

\begin{figure*}
\begin{center}
\epsfig{file=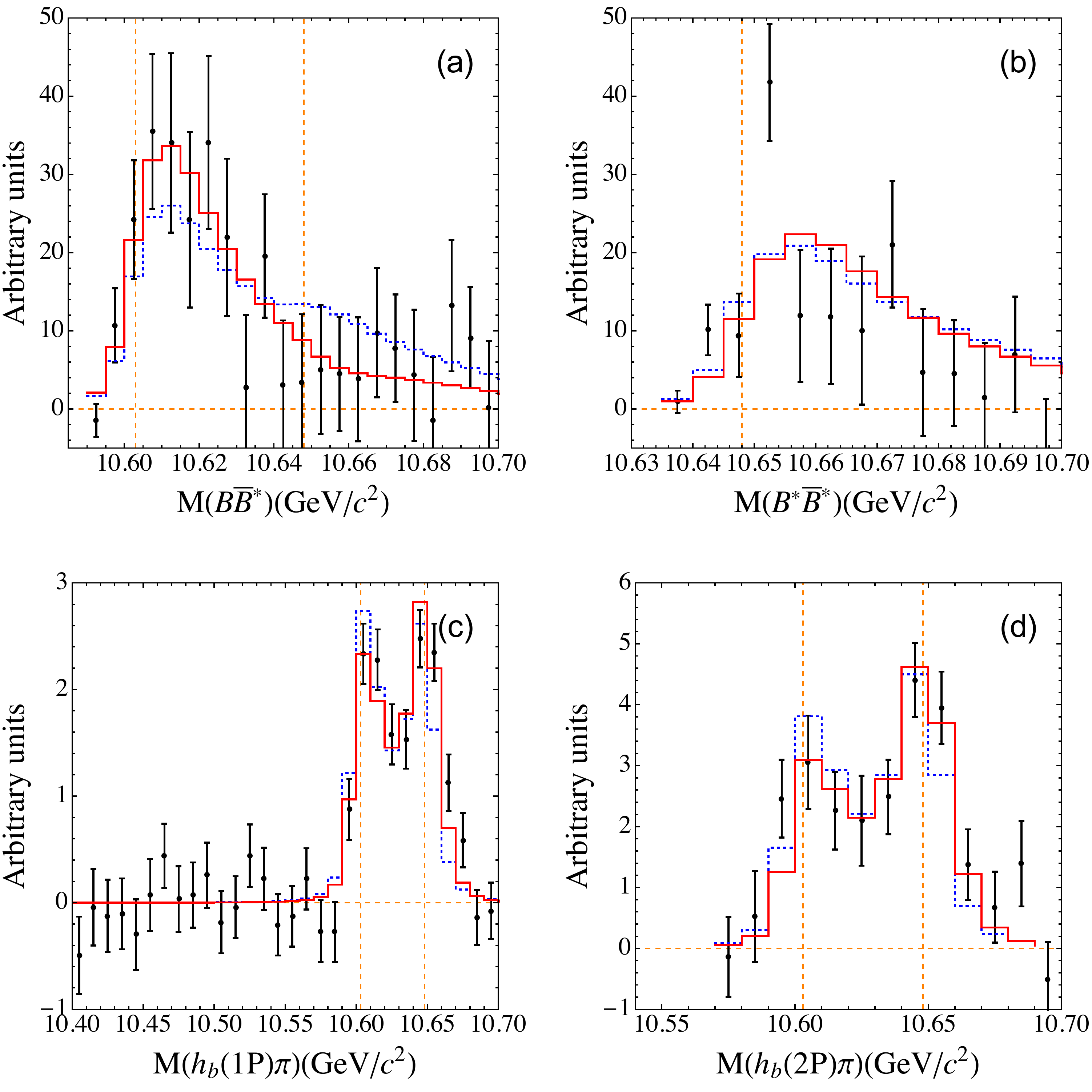, width=0.8\textwidth}
\caption{ \label{fig:fh} The fitted line shapes for Schemes A (blue dotted curves) and E (red solid curves)
in the elastic $B^{(*)}\bar{B}^*$ and in the inelastic $h_b(mP)\pi$ ($m=1,2$) channels. The experimental data are from Refs.~\cite{
Belle:2011aa,Garmash:2015rfd}.}
\end{center}
\end{figure*} 

We are now in a position to re-analyse the net effect from the OPE.
In Fig.~\ref{fig:fh}, we compare the results of the fits for Scheme E (red solid curves) with those where pions are switched off (Scheme A --- the blue dotted curves).
The quantitative improvement that one observes 
when proceeding from fit A to fit E is related to the residual dynamics from the OPE. Indeed, we checked that the $\mathcal{D}_{SD}$
contact term added to fit A does not improve the quality of the fit. 
Meanwhile, an attempt to improve the pionless fit A by adding all ${\cal O}(p^2)$ contact terms with no additional constraints 
fails: although the quality of the fit improves, 
the hierarchy of the resulting parameters is very unnatural, because 
the poles position in this fit is controlled by the NLO ${\cal O}(p^2)$ terms rather than by the LO ${\cal O}(1)$ ones in conflict
with the underlying assumption that higher derivative operators are suppressed.
For this reason this fit is not used in what follows. It is important to notice, however, that the EFT expansion is 
restored once pions are included (fits E, F and G). In particular, the transition from fit E to fits F or G, which corresponds to the addition of 
the ${\cal O}(p^2)$ $S$-$S$ contact terms, has a moderate (perturbative) impact on the line shapes as well as the poles.

\begin{figure*}
\begin{center}
\epsfig{file=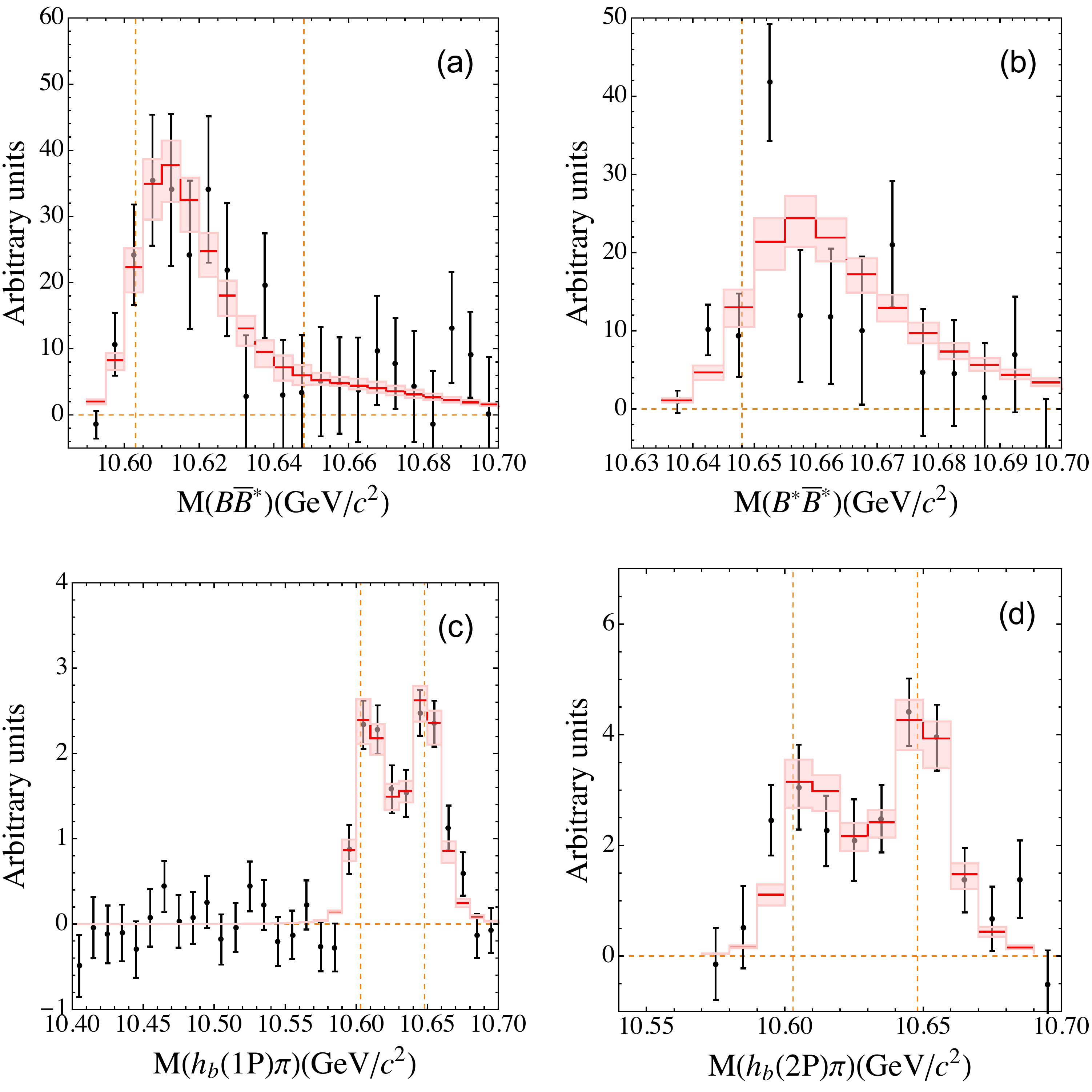, width=0.8\textwidth}
\caption{\label{fig:G} The fitted line shapes for Scheme G with the uncertainties corresponding to a $1\sigma$ deviation 
in the parameters of the fits. The experimental data are from Refs.~\cite{Belle:2011aa,Garmash:2015rfd}.}
\end{center}
\end{figure*}

Finally, in Fig.~\ref{fig:G} we present the results for Scheme G including the uncertainties which correspond to a $1\sigma$ deviation in the parameters including correlations.

In summary, the results presented in this section demonstrate that 
\begin{itemize}
\item[(i)] the data can be equally well described with or without the central $S$-wave 
OPE potential since the latter can be absorbed into a re-definition of the $S$-wave short-range contact terms; 
\item[(ii)] the inclusion of $D$ waves affects the 
line shapes noticeably which confirms the claims found in the literature that $D$ waves are important in the 
near-threshold charmonium-like and bottomonium-like systems \cite{Baru:2016iwj,Baru:2017gwo}. 
However, the current data call for the promotion of the $\mathcal{O}(p^2)$ $S$-wave-to-$D$-wave counter term $\mathcal{D}_{SD}$ to lower order
and for tuning this term to balance the $S$-$D$ dynamics from the OPE.
The residual effect from the OPE on the line shapes is, however, still sizeable:
\item[(iii)] the effect of the OEE interaction is negligibly small;
\item[(iv)] the data are essentially consistent with HQSS constraints imposed on the potential;
\item[(v)] no indication for the importance of $\mathcal{O}(p^2)$ contact interactions in the inelastic channels is seen in the data. 
\end{itemize}

\subsection{Dependence of the results on the regulator}
\label{sec:renorm}

\begin{figure*}
 \begin{center}
\epsfig{file=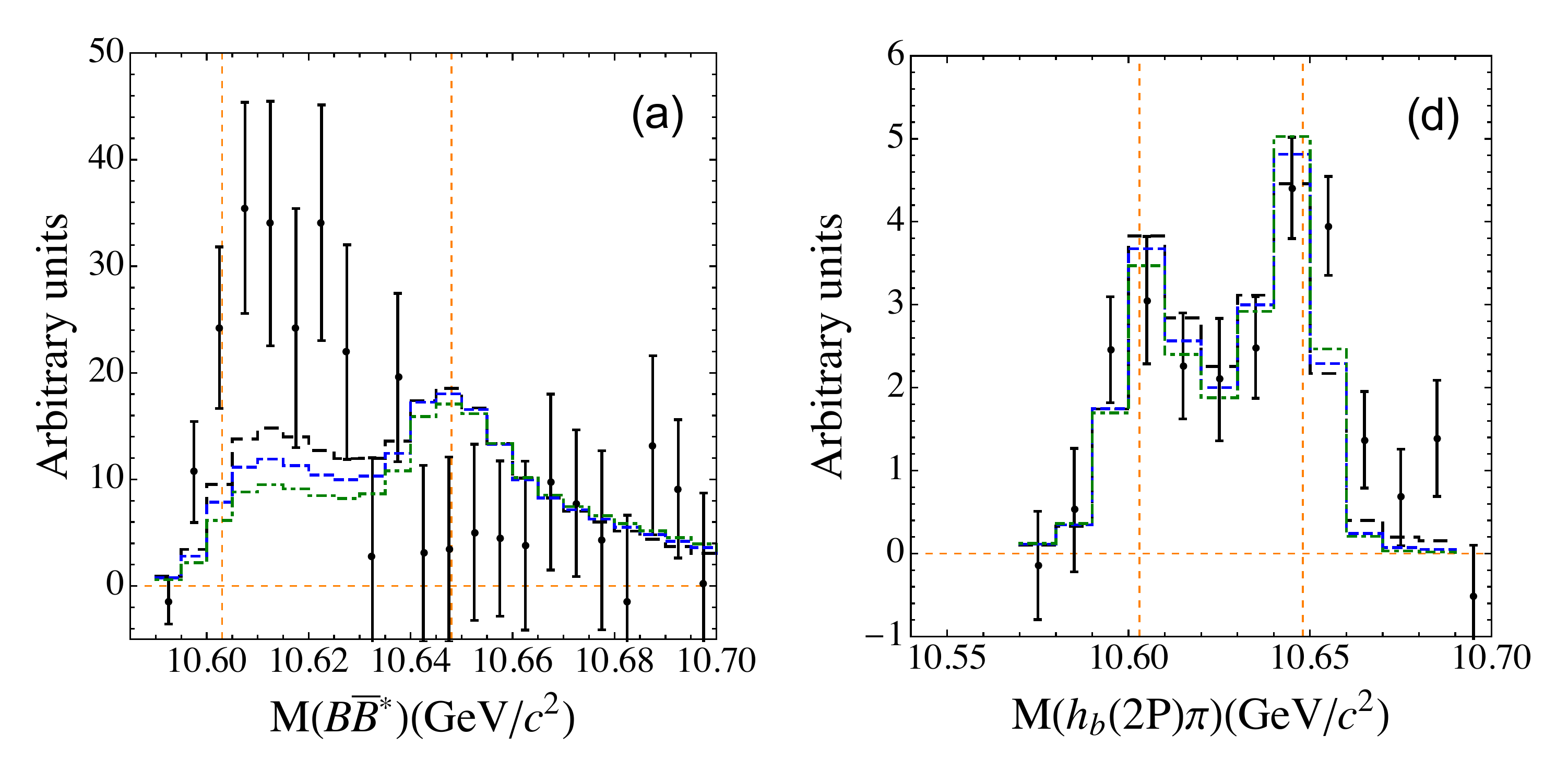, width=0.7\textwidth}
\hspace{0.6cm}
\epsfig{file=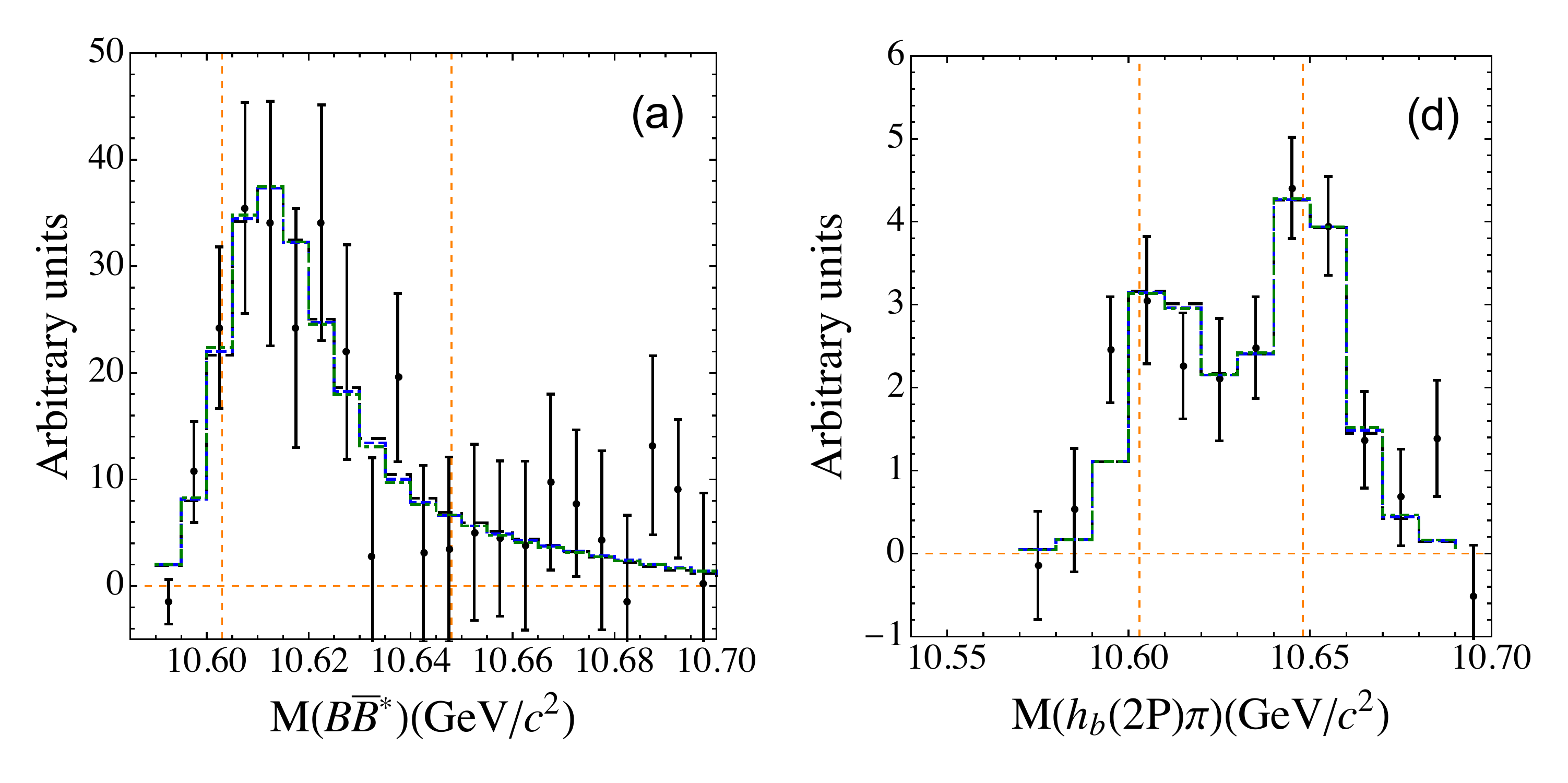, width=0.7\textwidth}
\caption{ \label{fig:CutF} 
The fitted line shapes for Scheme C (upper panel) and
Scheme G (lower panel)
in the elastic $B\bar{B}^*$ and inelastic $\pi h_b(2P)$ channels with sharp cut-offs 800~MeV (black long-dashed),
1000~MeV (blue dashed) and 1200~MeV (green dot-dashed), respectively. 
The experimental data are from Refs.~\cite{
Belle:2011aa,Garmash:2015rfd}.}
 \end{center}
\end{figure*}

In this chapter, we investigate if the promotion of the $S$-$D$ counter term is called for by the
renormalisation of the leading order amplitudes. 
In Fig.~\ref{fig:CutF} we show the regulator dependence of the 
results corresponding to the fits G and C, that is, those fits with and without the $\mathcal{O}(p^2)$ contact terms and OEE interactions in addition to the OPE. The
results for fit C show 
a clearly visible cut-off dependence in the $B\bar B^*$ and $h_b(2P)\pi$ channels
when the (sharp) cut-off in the LSEs, treated as a hard scale, is varied in a reasonable range 
from 800 to 1200 MeV (\emph{cf.} black long-dashed, blue dashed and green dotted-dashed curves).
Note that in an effective field theory where a potential, which is then iterated in a scattering equation, 
is expanded in terms of a given expansion parameter one cannot expect
a complete regulator independence of observable quantities at a given 
order. This implies that the regulator should not be chosen too large and that regulator effects of sub-leading order 
are common. It is, therefore, difficult to judge, if the mentioned
variations call for an additional counter term at LO or not. However, it is certainly interesting to observe that 
for the same cut-off variation the curves in fit G (as well as in fit F) remain unchanged indicating that the theory is fully renormalised. 
The same pattern is observed in the other channels which are therefore not shown.

\section{The poles position and the nature of the $Z_b$ states}\label{sec:poles}
 
\begin{figure*}
\epsfig{file=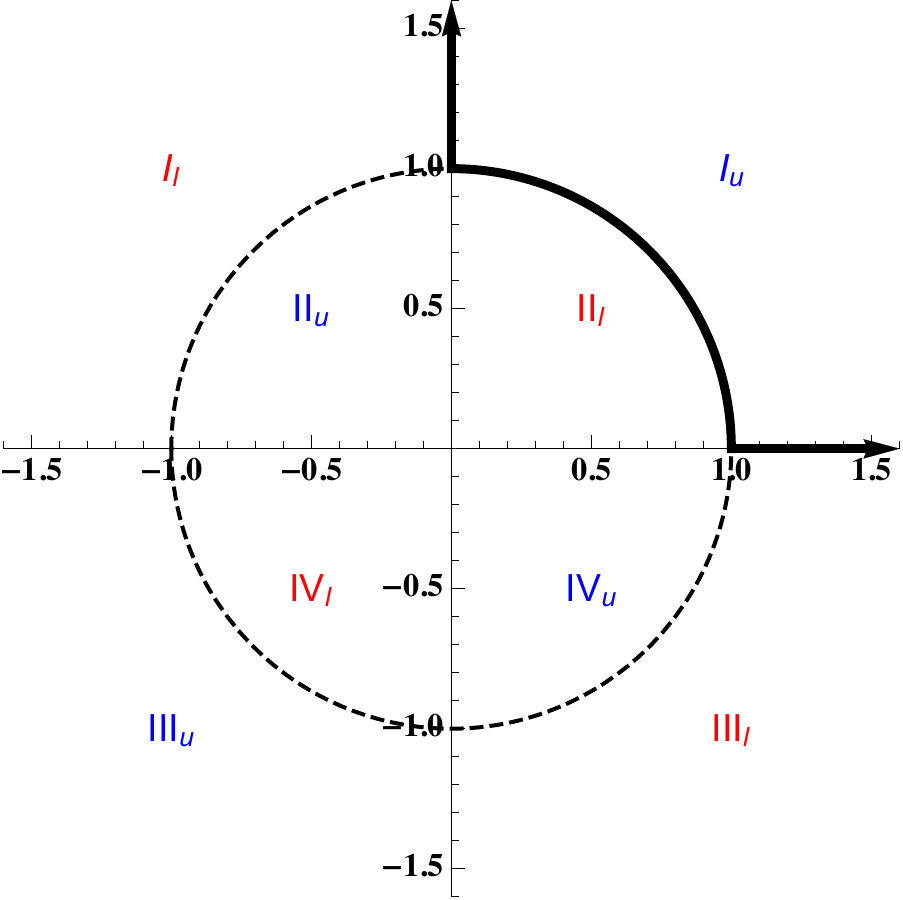, width=0.35\textwidth}
\hspace{1.3cm}
\epsfig{file=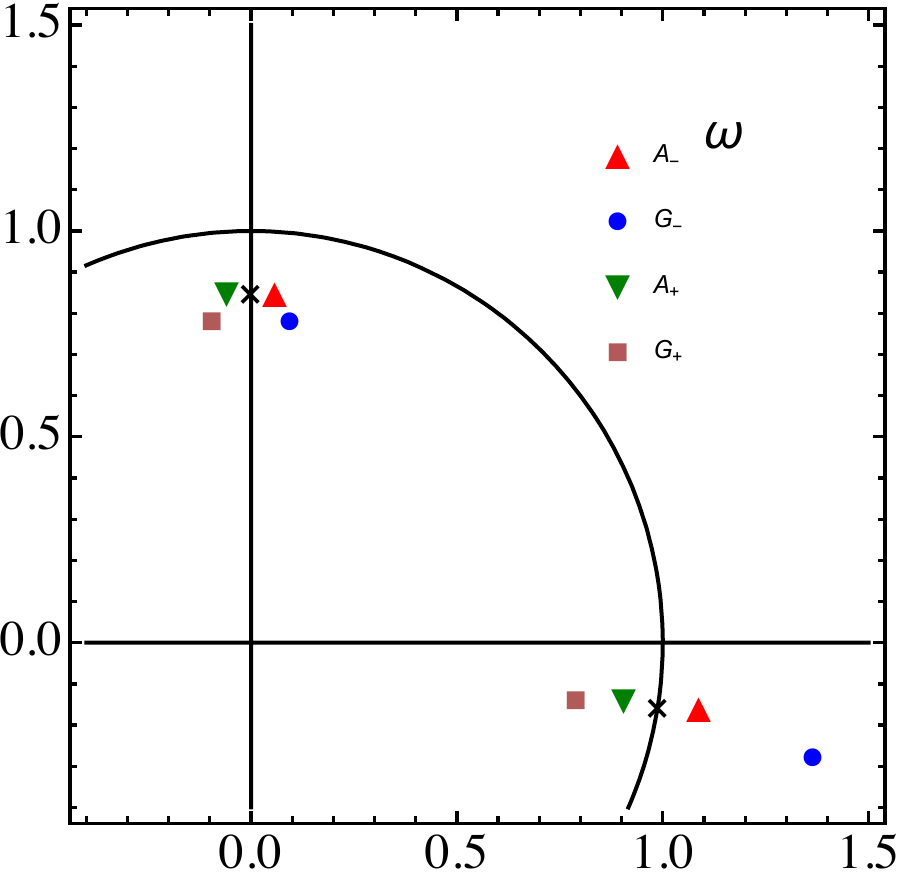, width=0.35\textwidth}\\
\caption{Left panel: The unitary-cut-free complex $\omega$-plane for the two elastic channels, $B\bar{B}^*$ and $B^*\bar{B}^*$, 
obtained from the four-Riemann-sheet complex energy plane by the conformal transformation (\ref{omegak1k2}). 
The eight regions separated by the unit circle and by the two axes correspond to the upper and lower half-planes [see the subscripts $u$ and $l$] 
in the four Riemann sheets of the energy plane denoted as RS-N with N=I, II, III, IV \cite{kato,Dudek:2016cru}. The bold line indicates the physical 
region of a real energy $E$ \cite{kato}. Right panel: The poles position in the complex $\omega$ plane for the fit Schemes A and G described in the text. 
Only the poles closest to the physical region are given while the distant poles are not shown.
The red triangle ($A_-$) and blue circle ($G_-$) stand for the poles for the fit Schemes A and G, respectively, 
with all the inelastic channels on their unphysical (${\rm Im}~k_{\rm in}<0$) Riemann sheets.
The green inverted triangle ($A_+$) and pink box ($G_+$) are for the poles for the fit Schemes A and G, respectively, 
with all the inelastic channels on their physical (${\rm Im}~k_{\rm in}>0$) Riemann sheets. The crosses stand for the poles for fit A when all inelastic channels are switched off.}
\label{fig:omega}
\end{figure*}

In this section we discuss the extraction of the poles of the amplitude in the complex plane. 
In general, in order to search for the poles in a multi-channel problem a multi-sheet Riemann surface in the complex energy plane needs to be invoked. 
However, we consider the four-sheet Riemann surface corresponding to the two elastic channels only, because all 
inelastic thresholds are remote and their impact on the poles of interest, which are located near the elastic thresholds, is expected to be minor. 
Then, for two coupled channels with the thresholds split by the mass difference $\delta$,
the four-sheeted Riemann surface can be mapped onto a single-sheeted plane of a variable which is traditionally denoted as
$\omega$ \cite{kato} and which, for a given energy $E=M-m_{B}-m_{B^*}$, is defined via
\bea
&\ds k_1=\sqrt{\frac{\mu_1\delta}2}\left(\omega+\frac1{\omega}\right),~
k_2=\sqrt{\frac{\mu_2\delta}2}\left(\omega-\frac1{\omega}\right),&\nonumber\\[-2mm]
\label{omegak1k2}\\[-2mm]
&\ds E=\frac{k_1^2}{2\mu_1}=\frac{k_2^2}{2\mu_2}+\delta=\frac{\delta}4\left(\omega^2+\frac1{\omega^2}+2\right),&\nonumber
\eea
where $\mu_1$ and $\mu_2$ are the reduced masses in the first ($B\bar{B}^*$) and in the second ($B^*\bar{B}^*$) elastic channel, respectively.
Then the one-to-one correspondence between the four Riemann sheets in the $E$-plane 
(denoted as RS-N, where N=I, II, III, IV) and various regions in the $\omega$-plane reads 
\begin{eqnarray*}
\mbox{RS-I}:&&\quad{\rm Im}~k_1>0,\quad{\rm Im}~k_2>0,\\
\mbox{RS-II}:&&\quad{\rm Im}~k_1<0,\quad{\rm Im}~k_2>0,\\
\mbox{RS-III}:&&\quad{\rm Im}~k_1<0,\quad{\rm Im}~k_2<0,\\
\mbox{RS-IV}:&&\quad {\rm Im}~k_1>0,\quad{\rm Im}~k_2<0.
\end{eqnarray*}
The corresponding regions in the $\omega$-plane are depicted in the first plot in Fig.~\ref{fig:omega}.
The thick solid line corresponds to the real values of the energy lying on RS-I.
It is easy to see that the physical region between the two thresholds 
corresponds to
$|\omega| = 1$, with both Re$(\omega)$ and Im$(\omega)$ positive,
and the thresholds at $E=0$ and $E=\delta$ are mapped to the points $\omega=\pm i$ and 
$\omega=\pm 1$, respectively. The inelastic channels in the energy plane are interpreted as one additional effective remote channel with the momentum $k_{\rm in}$ and, in order to find all relevant
poles, both possibilities with ${\rm Im}~k_{\rm in}>0$ and ${\rm Im}~k_{\rm in}<0$ are considered.\footnote{In the actual calculation, 
the various inelastic channels have different momenta. However, when searching for the poles, all inelastic channels are assumed to be on the same sheet, that is the imaginary parts of all 
inelastic momenta are synchronised to be either positive or negative.} As it is now effectively a three-channel problem, the number of Riemann sheets doubles and so does the number of poles 
representing the physics for each state. Hence, each pole has now its mirror partner located on a different sheet of the eight-sheeted Riemann surface but corresponding to the same physical state.

\begin{figure*}
\epsfig{file=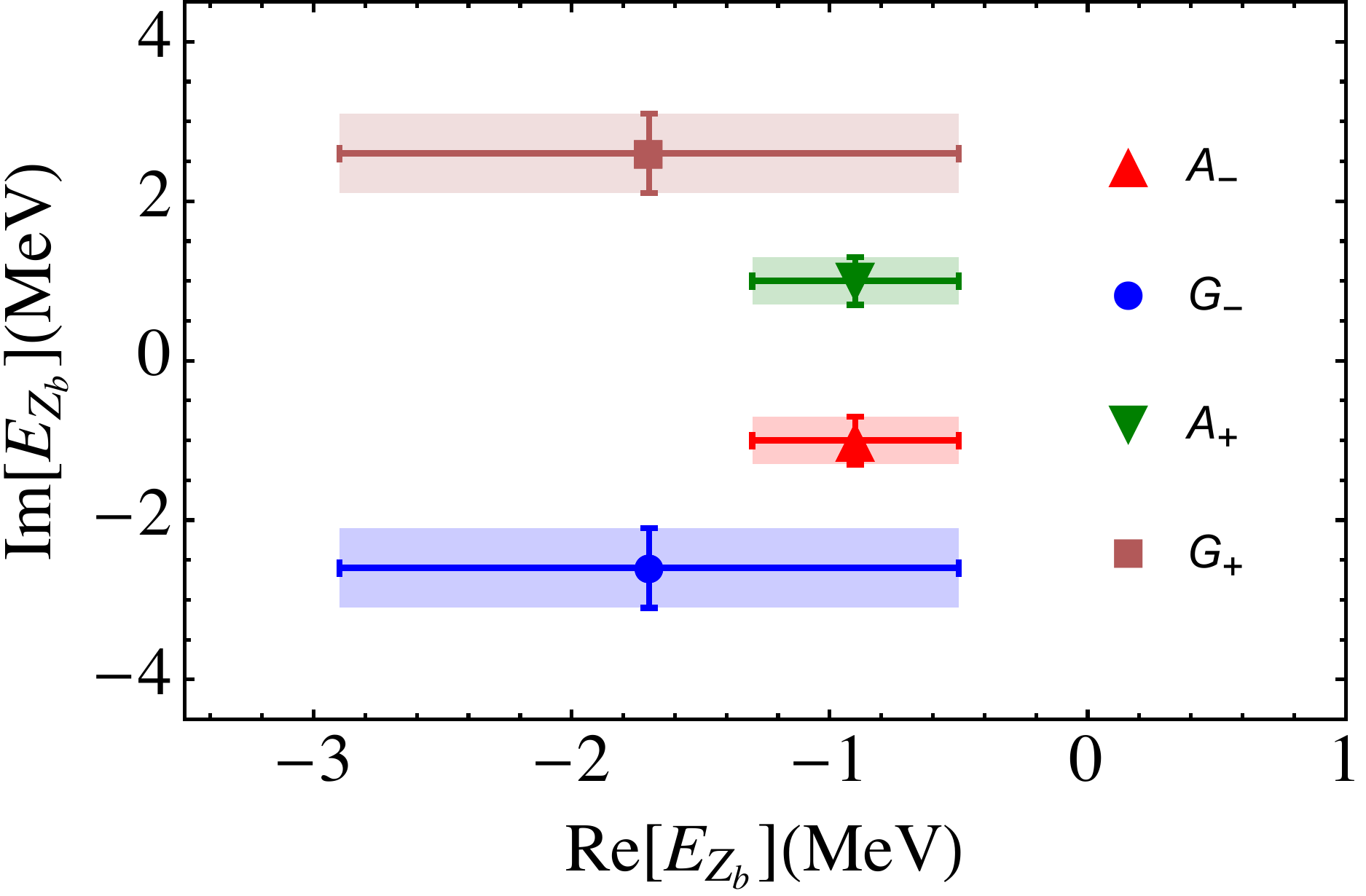, width=0.45\textwidth}\hspace{0.8cm}
\raisebox{-1.5mm}{\epsfig{file=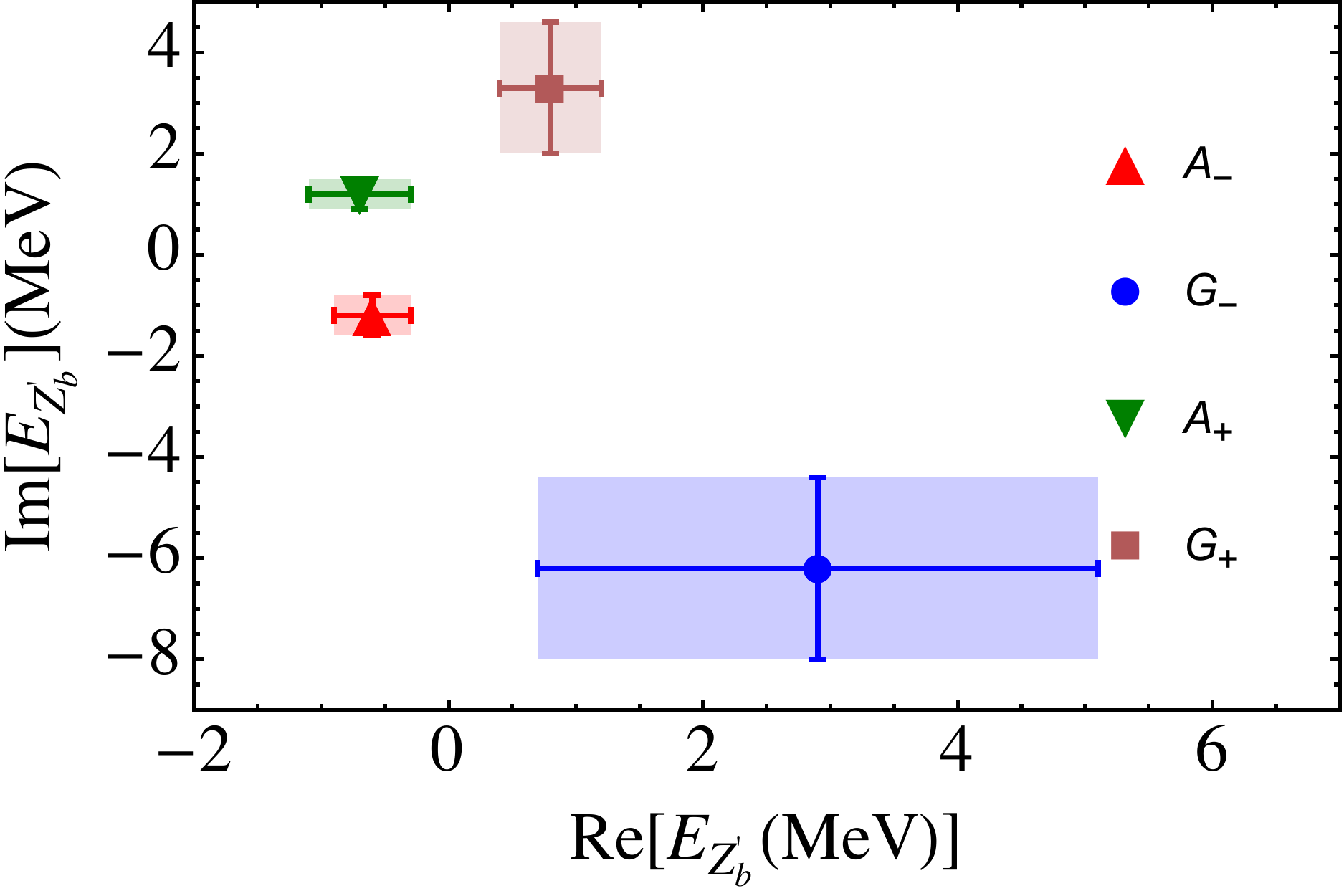, width=0.455\textwidth}}\\
\caption{The real and the imaginary part of the $Z_b$ (left panel) and $Z_b'$ (right panel) poles [see the definition in Eq.~(\ref{EZZ})] 
for fit Schemes A and G. For the notation of poles see Fig.~\ref{fig:omega}.}
\label{fig:energywidth}
\end{figure*}

The poles position in the $\omega$-plane for the fit Schemes A and G from the previous section is shown in the right panel of 
Fig.~\ref{fig:omega}. As long as the inelastic channels are switched off, one is back to the two-channel problem with one relevant pole corresponding to each state. For example, 
the poles for the fit Scheme A but without inelastic channels are shown by the crosses (x) in Fig.~\ref{fig:omega}: the cross on the imaginary axis on RS-II (close to 
$\omega =i$) corresponds to a virtual $B\bar B^*$ state pole associated with the $Z_b$ state while the other pole, residing on RS-IV (close to 
$\omega =1$), represents the physics related to the $Z_b'$ --- it is a virtual state in the $B^*\bar B^*$ channel slightly shifted to the complex plane due to coupled-channel
effects.\footnote{Both these poles have counterparts on RS-IV around $\omega =-i$ and $\omega =-1$, respectively, that is far away from the physical region. Analogous additional poles also 
appear for the other cases discussed below, but will not be mentioned anymore because they do not affect the line shapes.} When the effective inelastic channel is on, one arrives at a pair of poles 
for each state: one for ${\rm Im}~k_{\rm in}>0$ and one for ${\rm Im}~k_{\rm in}<0$ --- the corresponding solutions are labeled by ``$+$'' and ``$-$'', respectively. For the $Z_b$, the two resulting 
poles on 
RS-II$_+$ and RS-II$_-$ are symmetric with respect to the imaginary axis in the omega plane which results in complex-conjugate solutions for the energies (see Table \ref{tab:energy}). Unlike 
the $Z_b$, the poles for the $Z_b'$ on RS-III$_-$ and RS-IV$_+$ are slightly asymmetric due to coupled-channel effects. In any case, it is apparent that the poles obtained in the full calculation, 
including inelastic channels, reside in the vicinity of those found without inelastic channels, that implies that the role played by the inelastic channels is sub-leading in line with a molecular 
interpretation of the $Z_b$ states.

The corresponding energies evaluated relative to the relevant elastic threshold,
\be
E_{Z_b}=M_{Z_b}^{\rm pole}-m_B-m_{B^*},\quad E_{Z_b'}=M_{Z_b'}^{\rm pole}-2m_{B^*},
\label{EZZ}
\ee
are listed in Table~\ref{tab:energy} and are visualised in Figs.~\ref{fig:omega} and \ref{fig:energywidth} 
(the errors in the poles position corresponds to a $1\sigma$ deviation for the whole parameter list). 
Note that the sign convention is such that positive energies refer to above-thresholds poles --- see definition (\ref{EZZ}).

\begin{table*}
\begin{center}
\begin{tabular}{|c|c|c|}
\hline 
scheme & $E_{Z_{b}}(\mathrm{MeV})$ & $E_{Z_{b}^{\prime}}(\mathrm{MeV})$\tabularnewline
\hline 
\hline 
$A_-$ & $(-0.9\pm 0.4)-i(1.0\pm 0.3)$ & $(-0.6\pm 0.3)-i(1.2\pm0.4)$\tabularnewline
\hline
$G_-$ & $(-1.7\pm 1.2)-i(2.6\pm 0.5)$ & $(2.9\pm 2.3)-i(6.2\pm 1.8)$\tabularnewline
\hline 
$A_+$ & $(-0.9\pm 0.4)+i(1.0\pm 0.3)$ & $(-0.7\pm 0.4)+i(1.2\pm 0.3)$\tabularnewline
\hline
$G_+ $ & $(-1.7\pm 1.2)+i(2.6\pm 0.5)$ & $(0.8\pm 0.4)+i(3.3\pm 1.3)$ \tabularnewline
\hline 
\end{tabular}
\end{center}
\caption{The energies $E_{Z_b}$ and $E_{Z_b'}$ (see the definition in Eq.~(\ref{EZZ})) for the fit Schemes A and G. The energies denoted as $A_-$ and $G_-$ stand for 
the poles for the fit Schemes A 
and G, respectively, 
with all the inelastic channels on their unphysical (${\rm Im}~k_{\rm in}<0$) Riemann sheets.
The energies denoted as $A_+$ and $G_+$ are for the poles for the fit Schemes A and G, respectively, 
with all the inelastic channels on their physical (${\rm Im}~k_{\rm in}>0$) Riemann sheets. 
The errors correspond to a $1\sigma$ deviation 
in the fitted parameters.}
\label{tab:energy}
\end{table*}

The two poles for Fit A representing the results based on S-wave contact interactions (shown as an up- (red) and down-pointing (green) triangles in Figs.~\ref{fig:omega} and \ref{fig:energywidth}) 
are essentially consistent
with those obtained in Ref.~\cite{Guo:2016bjq} 
--- see Fig.~8 of the quoted paper. 
As one can see from Table~\ref{tab:energy} and from Figs.~\ref{fig:omega} and \ref{fig:energywidth}, 
the inclusion of the OEE and especially the OPE and ${\cal O}(p^2)$ contact interactions in Fit G changes the poles position to some extent but all the poles 
reside in the vicinity of the corresponding thresholds. 
This result is consistent with the expectation that the line shapes are controlled
predominantly by the poles position.
Indeed, although the parameters of fits A and one of our best fits (fit G) are very different (see Tables~\ref{tab:par-1} and \ref{tab:par-2}), fit G provides a better but still comparable 
description of the data as 
fit A, as one can see from the line shapes shown in Fig.~\ref{fig:fh} and from the values of $\chi^2/\mbox{d.o.f.}$ quoted in 
Table~\ref{tab:par-1}. Thus, both fits describe the $Z_b$ state as a shallow virtual state located below the $B\bar B^*$ threshold. 
Meanwhile, the $Z_b'$ state is consistent with both a virtual state and an above-threshold resonance interpretation, with the latter option preferred by the pionful fits (like fit G). 

\section{Summary}\label{sec:sum}

This work continues a series of papers aimed at a systematic description of the line shapes of near-threshold resonances in general and, 
in particular, at understanding the nature and properties of the $\zb$ and $\zbp$ states in the spectrum of bottomonium. Unlike 
the previous papers in the series \cite{Hanhart:2015cua,Guo:2016bjq}, in this work, we do not resort to an analytic parameterisation for the line shapes but 
rely on an EFT approach to calculate the line shapes explicitly. 
In particular, in order to obtain the production and scattering amplitudes, we construct the effective potential to leading order in the chiral and heavy
quark expansion
and iterate it to all orders employing coupled-channel
Lippmann-Schwinger equations. This allows us to verify the accuracy of the practical parameterisation suggested in Ref.~\cite{Hanhart:2015cua}
and to estimate the role played by the $\pi$- and $\eta$-exchanges, which cannot be straightforwardly incorporated into the 
scheme of Refs.~\cite{Hanhart:2015cua,Guo:2016bjq} because of their non-separable form. 

The results of this work can be summarised as follows:
\begin{itemize}
\item We find that 
the distributions obtained from the direct numerical solution of the LSEs with just S-wave contact potentials 
provide a nearly identical description of the data to that achieved using the parameterisation derived in Refs.~\cite{Guo:2016bjq,Hanhart:2015cua,Hanhart:2016eyl}.
\item We include the OPE interaction on top of the contact potentials in the elastic channels and demonstrate that 
the inclusion of the $S$-wave OPE affects the line shapes only marginally. 
After a re-fit to the data required to appropriately renormalise the short-range interactions,
basically the whole $S$-wave OPE contribution can be absorbed to a re-definition of the $S$-wave contact interactions.
 Therefore, we do not support the claim of Ref.~\cite{Voloshin:2015ypa} that OPE changes the line shape of near threshold states by about 30\%.
\item For the actual value of the splitting of the two elastic thresholds, $\delta\approx 45$~MeV, the momentum scale $p_{\rm typ}\simeq 500$ MeV which controls the coupled-channel ($B\bar 
B^*$-$B^*\bar B^*$)
dynamics is relatively large --- see Eq.~(\ref{ptyp}). For such momenta, 
the role played by the OPE in $D$ waves turns out to be significant even after a re-fit, so that one cannot neglect 
this effect in order to extract the resonance parameters with a sufficiently high accuracy. 
However, the resulting significant distortion of the line shapes from the OPE is not supported by the currently available experimental data in the $B\bar B^*$ channel: Indeed, 
a clear bump structure around the $B^*\bar B^*$ threshold unavoidably generated by the $S$-$D$ OPE transitions is not seen in the most recent data. 
In order to cure this, we include the additional contact term allowed by heavy quark symmetry and chiral symmetry at order $\mathcal{O}(p^2)$ to find that 
the resulting line shapes are in a very good agreement with the data, with a reduced $\chi^{2}/\text{d.o.f.}$ around unity. 
The role played by various $\mathcal{O}(p^2)$ contact interactions is different: on the one hand, the single contact term which contributes to the 
$S$-to-$D$-wave transitions 
absorbs a large part of the $S$-$D$ OPE piece and, in addition, brings an additional residual contribution --- both effects together
improve the quality of the fit considerably; on the other hand, two allowed $S$-$S$ contact interactions play 
a sub-leading role resulting only in a marginal change in the fits. Finally, we conclude that after the inclusion of the 
$\mathcal{O}(p^2)$ contact interactions the residual effect from the OPE moderate.
\item There are indications that the inclusion of the $S$-$D$ contact term at leading order is required by 
renormalisation. However, to make this statement more sound a complete calculation to next-to-leading order would be necessary.
Since this would come with additional parameters that need to be fixed by data, we postpone this effort until improved data
become available. 
\item We find that the data are consistent with HQSS symmetry constraints imposed on the potential.
\item The data do not call for the inclusion of $\mathcal{O}(p^2)$ contact interactions in the elastic-to-inelastic 
transitions.
 \item The effect from the $\eta$-exchange potential is negligible. 
\item We extract the position of the poles responsible for the $\zb$ and $\zbp$ and find them to reside on the unphysical Riemann sheets just below ($Z_b$) or just above ($Z_b'$) the corresponding 
elastic threshold.
\end{itemize}

Before closing, we would like to stress that the observed strong cancellation between the OPE and the additional ${\cal O}(p^2)$ $S$-$D$ contact terms 
is very puzzling. The pion plays a very special role in QCD due to its intimate connection to the spontaneous breaking of chiral symmetry. In addition, since the $D^*D\pi$ coupling is known (see 
Eq.~(\ref{gb})), the OPE comes without adjustable parameters. Still, data demand 
a very large cancellation of this part of the potential. 
Whether or not this cancellation is accidental in the system 
at hand or if it has deeper reasons in QCD calls for further studies which, however, go beyond the scope of this work. Clearly, more accurate experimental data for the $Z_b$'s, especially in the 
elastic channels, and, hopefully, new data for their spin partner states would be of great relevance and importance for such studies.

In conclusion, the results presented provide a good understanding of the line shapes with the $\chi^2$/d.o.f. $\simeq 1$ for the best fit. Given stability of the results to various higher-order 
interactions included and the quality of the fits, it is unlikely that 
higher-order contributions not included explicitly in the current study, such as two-pion exchange contributions at next-to-leading order, affect the conclusions of the analysis. 
As a consequence, the LECs extracted from the presented fits can be used to make predictions for the molecular spin-partner states within the same theoretical framework and without introducing any 
additional parameters \cite{spinpartners}. It remains to be seen if the 
strong cancellation of the one-pion exchange and the short-range contributions, which takes place for the $Z_b$'s, 
persists also for their spin partners. 

\begin{acknowledgments}
The authors are grateful to Michael D\"oring, Evgeny 
Epelbaum, Feng-Kun Guo, Zhi-Hui Guo, Roman Mizuk, Deborah R\"onchen and Akaki Rusetsky for fruitful and enlightening discussions and comments. 
This work was supported in part by the DFG (Grant No. TRR110) and the NSFC (Grant No. 11621131001) through the
funds provided to the Sino-German CRC 110 ``Symmetries and the Emergence of Structure
in QCD''. Work of V.B. and A.N. was supported by the Russian Science Foundation (Grant No. 18-12-00226).
\end{acknowledgments}

\appendix

\section{Effective Lagrangians}\label{app:Lag}
The effective Lagrangian describing isovector $B^{(*)} \bar{B}^{(*)}$ scattering at low energies reads
\begin{widetext}
\bea
&&{\cal L}={\rm Tr}\left[H^\dagger_a \left(i \partial_0 +\frac{\vec\nabla^2}{2 M}\right)_{ba} H_b\right]+\frac{\Delta}{4}{\rm Tr}[H^\dagger_a\sigma^i H_a \sigma^i]\nn \\
&&+{\rm Tr}\left[\Hb^\dagger_a\left(i\partial_0+\frac{\vec\nabla^2}{2M}\right)_{ab}\Hb_b\right]+\frac{\Delta}{4}{\rm Tr}[\Hb^\dagger_a\sigma^i\Hb_a \sigma^i]\nn \\
&&-\frac{C_{10}}{8}{\rm Tr}[\bar{H}^\dagger_a\tau^A_{aa^\prime}H^\dagger_{a^\prime}H_b\tau_{bb^\prime}^A\bar{H}_{b^\prime}] 
-\frac{C_{11}}{8}{\rm Tr}[\bar{H}^\dagger_a\tau^A_{aa^\prime}\sigma^i H^\dagger_{a^\prime}H_b\tau_{bb^\prime}^A\sigma^i\bar{H}_{b^\prime}]\nn\\
&&-\frac{D_{10}}8\left\{{\rm Tr}[\nabla^i\bar{H}^\dagger_a\tau^A_{aa^\prime}\nabla^i H^\dagger_{a^\prime}H_b\tau_{bb^\prime}^A\bar{H}_{b^\prime} 
+{\rm Tr}[\bar{H}^\dagger_a\tau^A_{aa^\prime}H^\dagger_{a^\prime}\nabla^i H_b\tau_{bb^\prime}^A\nabla^i\bar{H}_{b^\prime}]\right\} \label{Lag}\\
&&-\frac{D_{11}}{8}\left\{{\rm Tr}[\nabla^i\bar{H}^\dagger_a\tau^A_{aa^\prime}\sigma^j\nabla^i H^\dagger_{a^\prime}H_b\tau_{bb^\prime}^A\sigma^j \bar{H}_{b^\prime}+
{\rm Tr}[\bar{H}^\dagger_a\tau^A_{aa^\prime}\sigma^j H^\dagger_{a^\prime}\nabla^i H_b\tau_{bb^\prime}^A\sigma^j\nabla^i\bar{H}_{b^\prime}]\right\}\nn\\
&&-\frac{D_{12}}{8}\left\{{\rm Tr}[(\nabla^i\bar{H}^\dagger_a \tau^A_{aa^\prime}\sigma^i \nabla^j H^\dagger_{a^\prime}+
\nabla^j\bar{H}^\dagger_a \tau^A_{aa^\prime}\sigma^i\nabla^i H^\dagger_{a^\prime}-\frac23\delta^{ij}\nabla^k\bar{H}^\dagger_a\tau^A_{aa^\prime}\sigma^i \nabla^k H^\dagger_{a^\prime})H_b
\tau_{bb^\prime}^A \sigma^j \bar{H}_{b^\prime}]\right. \nn \\
&&+\left.{\rm Tr}[\bar{H}^\dagger_a \tau^A_{aa^\prime} \sigma^i H^\dagger_{a^\prime} (
 \nabla^i H_b \tau_{bb^\prime}^A \sigma^j \nabla^j \bar{H}_{b^\prime} 
+\nabla^j H_b \tau_{bb^\prime}^A \sigma^j \nabla^i \bar{H}_{b^\prime} 
- \frac23\delta^{ij} \nabla^k H_b \tau_{bb^\prime}^A \sigma^j \nabla^k \bar{H}_{b^\prime} ) ] 
\right \}\nn,
 \eea
 \end{widetext}
where the terms in the first row in Eq.~\eqref{Lag} stand for the leading heavy hadron chiral perturbation theory Lagrangian of Refs. \cite{Wise:1992hn,Burdman:1992gh,Yan:1992gz},
written in the two-component notation of Ref.~\cite{Hu:2005gf}. The terms in the second row represent the Lagrangian for anti-heavy mesons. The heavy mesons and anti-heavy 
mesons interact via the remaining terms in the Lagrangian: the third row in Eq.~\eqref{Lag} corresponds to $O(p^0)$ $S$-wave contact interactions~\cite{Mehen:2011yh,AlFiky:2005jd}, 
while the last three rows represent the ${\cal O}(p^2)$ contact terms with two derivatives. The contact terms $\propto D_{10}$ and $D_{11}$ contribute to 
$S$-wave interactions 
while the last term $\propto D_{12}$ is projected out to give rise to the $S$-$D$ transitions. 
Since we are only interested in the $S$-$S$ and $S$-$D$ transitions for the $B^{(*)} \bar{B}^{(*)}$ scattering, all the terms of the kind $\propto \nabla^i H^\dagger \nabla^j H$ 
contributing to $P$ waves were dropped. We also note here that the expansion in spatial derivatives employed in the Lagrangian (\ref{Lag}) yields contributions to four-point vertices rather 
than to two-point vertices and propagators. This expansion 
is controlled by the scale provided by the range of forces
(for example, by the mass of the lightest $t$-channel exchange particle and not by the heavy-quark mass),
so that arbitrary coefficients $D$ in front of the terms with derivatives do not violate the reparametrisation invariance discussed in detail in Ref.~\cite{Luke:1992cs}.

In Eq.~\eqref{Lag} $H_a = P_a+ V_a^i \sigma^i$ stands for the heavy meson superfield combining a pseudoscalar ($P_a$) and a vector ($V_a$) meson, $a$ and $b$ are $SU(2)$ isospin indices, and the 
isospin matrices are normalised via the trace as $\tau^A_{ab} \tau^B_{ba} =2\delta^{AB}$. 

The charge and Hermitian conjugate operators for the superfield $H_a$ read
\begin{widetext}
\be
\bar{H}_a=\sigma_2\mathcal{C}H_a^T\mathcal{C}^{-1}\sigma_2=\bar{P}_a-\bar{V}_a^i\sigma^i,\quad
H_a^\dagger=P_a^\dagger+V_a^{i\dagger}\sigma^i,\quad
\bar{H}_a^\dagger=\bar{P}_a^\dagger-\bar{V}_a^{i\dagger}\sigma^i,
\ee
and they transform as
\begin{eqnarray}
H_a\stackrel{\mathcal{P}}{\longrightarrow}-H_a,\quad H_a\stackrel{\mathcal{C}}{\rightarrow}\sigma_2\bar{H}_a^T\sigma_2,\quad 
H_a\stackrel{\mathcal{S}}{\rightarrow}SH_a,\quad H_a\stackrel{\mathcal{U}}{\rightarrow}H_bU_{ba}^{\dagger},\nonumber \\
\quad\bar{H}_a\stackrel{\mathcal{P}}{\longrightarrow}-\bar{H}_a,\quad\bar{H}_a\stackrel{\mathcal{C}}{\rightarrow}\sigma_2H_a^T\sigma_2,\quad\bar{H}_a
\stackrel{\mathcal{S}}{\rightarrow}\bar{H}_a\bar{S}^{\dagger},\quad\bar{H}_a\stackrel{\mathcal{U}}{\rightarrow}U_{ab}\bar{H}_b
\end{eqnarray}
\end{widetext}
under the parity ($\mathcal{P}$), charge ($\mathcal{C}$), heavy-quark spin ($\mathcal{S}$), and chiral ($\mathcal{U}$) transformation. 
The isoscalar contributions in Lagrangian~\eqref{Lag} can be easily restored if one makes a replacement of the isospin Pauli matrices by the corresponding 
Kronecker delta symbols, that is, for example, $\tau_{ab} \to \delta_{ab}$.

Deriving the Feynman rules from the interaction part of Lagrangian (\ref{Lag}) one readily obtains the effective potential given in Eq.~\eqref{vfull}, where 
\bea
&\mathcal{C}_d=\ds -C_{11}-C_{10}, \quad \mathcal{C}_f= C_{10}-C_{11},& \quad \\ \nonumber
&\mathcal{D}_d=\ds -D_{11}-D_{10}, \quad \mathcal{D}_f= D_{10}-D_{11}, &\\ &\mathcal{D}_{SD}=\ds \frac{2\sqrt{2}}{3}D_{12}.&\nonumber
\eea

\section{Scaling of higher partial waves in elastic-to-inelastic transitions}
\label{app:inelD}

\begin{figure}
\centerline{\epsfig{file=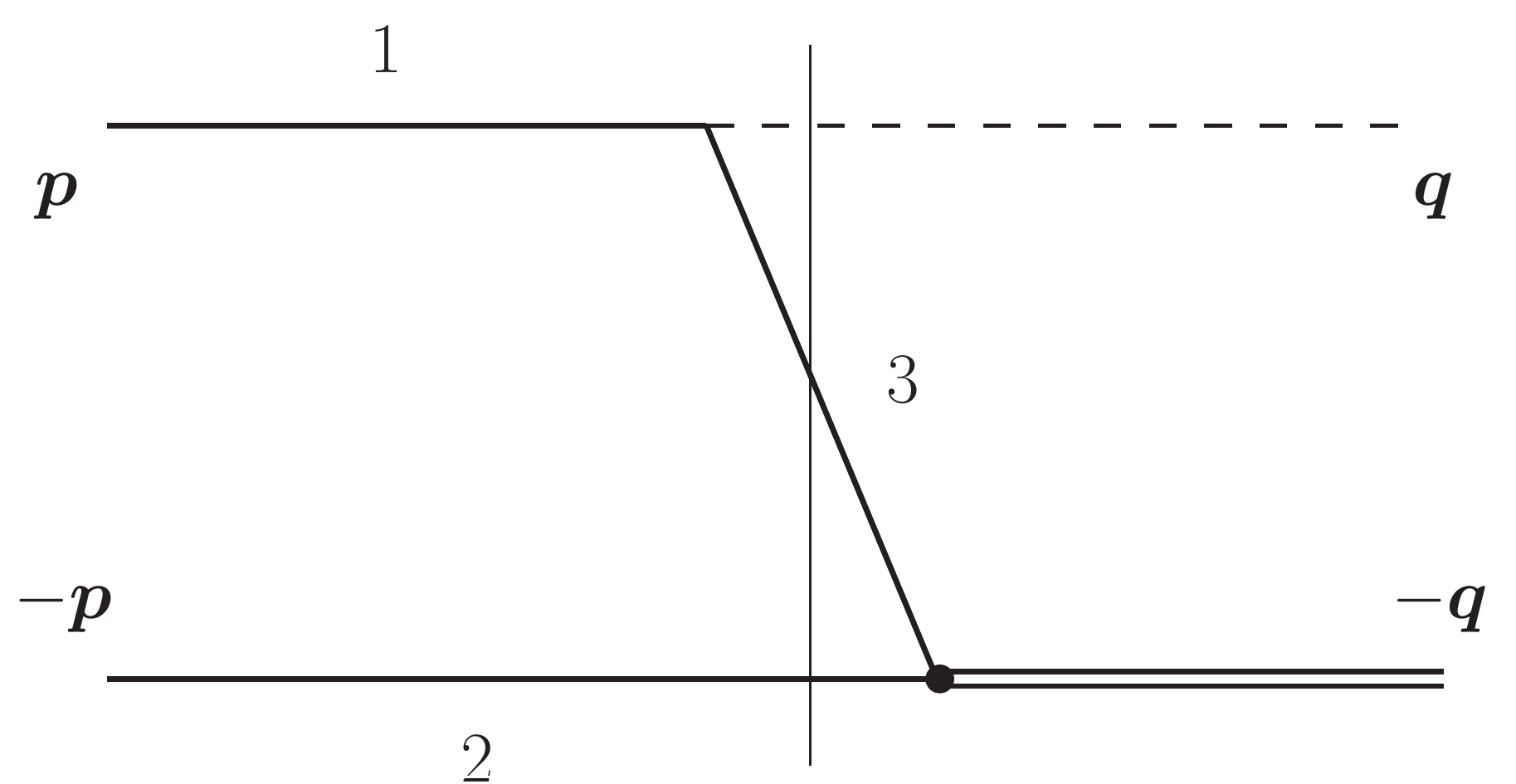, width=0.35\textwidth}}
\caption{Diagram underlying the transitions from an
elastic to an inelastic channel. Solid, dashed and double line denote $B^{(*)}$ mesons, the pion
and a heavy quarkonium, respectively. The thin vertical line indicates the time slice of
relevance for the discussion.}\label{fig:inelD}
\end{figure} 

To understand a possible
role of the $D$ waves in the transitions from an elastic to an inelastic channel one needs to study 
the diagram shown in Fig.~\ref{fig:inelD}. Taken as a time-ordered diagram it gives for the 
intermediate state pinpointed by the thin vertical line
\begin{widetext}
\begin{equation}
G^{-1}_{\rm trans.}(M,\vep,\vec q)=E_\pi(q)+\left(m_3+\frac{(\vep-\vec q)^2}{2m_3}\right)+\left(m_2+\frac{\vep^2}{2m_2}\right)-M.
\end{equation}
\end{widetext}
Since the energy region of interest is located near the $Z_b$'s states, it is sufficient for the argument to estimate the on-shell transition potential which calls for the substitution
$M=m_1+m_2+\vep^2/(2\mu)$, where $\mu=m_1m_2/(m_1+m_2)$ denotes the reduced mass. As an example (but without loss of generality for the argument), we now
set $m_1=m_{B^*}$ and $m_2=m_B$ that implies $m_3=m_B$ and $p$ ranges from 0, at the $B\bar B^*$ threshold, to 
$p\sim p_{\rm typ}$ (defined in Eq.~(\ref{ptyp})), when the $B^*\bar B^*$ threshold is approached.
Then we get
\begin{equation}
G^{-1}_{\rm trans.}(M,\vep,\vec q)=E_\pi(q)+ \frac{pq}{m_B}x+\frac{q^2}{2m_B}-\delta,
\end{equation}
where $x=\cos(\hat{\vep\veq})$ and tiny corrections $\propto \delta\, p^2_{\rm typ}/2m_B$ were neglected. It is the term $\propto x$ that eventually supports
higher partial waves in the elastic channel. Then the $D$ waves are suppressed relative to the $S$ waves
by the factor 
\be
\left(\frac{p_{\rm typ}}{m_B}\right)^2\left(\frac{q}{E_\pi(q)+q^2/(2m_B)-\delta}\right)^2.
\label{DSratio}
\ee

The values of $q$ range from about 200 MeV, for the transition to the $\Upsilon(3S)\pi$ channel, to 1.1 GeV, for the transition to the $\Upsilon(1S)\pi$ channel. It is easy to verify that, for such
values of $q$, the second factor in Eq.~(\ref{DSratio}) is close to unity, that
provides a justification for the estimate used in the main text --- see the discussion above Eq.~(\ref{via}).

\section{Flavour symmetry and the pseudoscalar exchange potentials}\label{app:eta}

In this appendix we discuss the pseudoscalar exchange potential between the heavy mesons. For definiteness, we stick to the 
$B\bar{B}^*$ channel as to an example. The flavour projector for a pseudoscalar ($P$ or $\bar P$) and a vector ($\bar V$ or $V$) for the $1^{+-}$ quantum numbers reads
\begin{eqnarray}
\mathcal{P}_{B\bar{B^{*}}}(1^{+-})=\frac12\left (P\tau^{A}\bar{V}-V\tau^{A}\bar{P}\right),
\end{eqnarray} 
where the overall factor 1/2 ensures the proper normalisation of the projectors,
\begin{widetext}
\begin{eqnarray}
\mathcal{P}_{B\bar{B^{*}}}(1^{+-})^{\dagger}\mathcal{P}_{B\bar{B^{*}}}(1^{+-})&=&\frac14{\rm Tr}\left[\left
(\bar{V}^{\dagger}\tau^{A}{P}^{\dagger}-\bar{P}^{\dagger}\tau^{A}V^{\dagger}\right )\left (P\tau^{B}\bar{V}-V\tau^{B}\bar{P}\right )\right]=\frac14{\rm Tr}\left( 2 \tau^A \tau^B 
\right)= \delta^{AB}.\nonumber
\end{eqnarray}

Here, like in Appendix~\ref{app:Lag}, the standard normalisation for the isospin matrices was used, ${\rm Tr}[\tau^A\tau^B]=2\delta^{AB}$.

Then, the flavour factor involving isospin and C-parity for the OPE diagram can be evaluated as
\begin{eqnarray}\nonumber
&&\frac14{\rm Tr}\left[ \left(
\bar{V}^{\dagger}\tau^{A}{P}^{\dagger}-\bar{P}^{\dagger}\tau^{A}V^{\dagger}\right ) 
\left( P^{\dagger}\tau^{a}V \right )
\left(
V\tau^{B}\bar{P}-P\tau^{B}\bar{V}
\right)
\left(
\bar{V}^{\dagger}\tau^{a}\bar{P}
\right)\right]\\ 
&+&\frac14{\rm Tr}\left[ \left(
\bar{V}^{\dagger}\tau^{A}{P}^{\dagger}-\bar{P}^{\dagger}\tau^{A}V^{\dagger}\right ) 
\left( V^{\dagger}\tau^{a}P \right )
\left(
V\tau^{B}\bar{P}-P\tau^{B}\bar{V}
\right)
\left(
\bar{P}^{\dagger}\tau^{a}\bar{V}
\right)\right]
\nonumber\\[-2mm]
\label{ea:appE}\\[-2mm]
&=& \frac14 {\rm Tr}\left(- \tau^{A}\tau^{a}\tau^{B}\tau^{a}- 
\tau^{A}\tau^{a}\tau^{B}\tau^{a} 
\right) =\delta^{AB},
\nonumber
\end{eqnarray} 
\end{widetext}
where
at the last step above an easily verified relation 
\be
{\rm Tr}\left[\tau^{A}\tau^{a}\tau^{B}\tau^{a}\right]=-2\delta^{AB}
\label{tracepion}
\ee
was used. It is this part of the calculation which makes difference between the OPE and OEE potentials. Indeed, in the latter case the 
trace takes the form 
\be
{\rm Tr}\left[\tau^{A}\hat1\tau^{B}\hat1\right]=2\delta^{AB},
\label{traceeta}
\ee
where the unit matrices correspond to the $\eta$ emission/absorption vertices which substitute the Pauli matrices from the pion 
vertices --- see Eq.~(\ref{tracepion}).

Finally, considering an additional factor $1/\sqrt3$ which comes from the 8-th Gell-Mann matrix for the $SU(3)$ group --- see Eq.~(\ref{pieta}) --- and which,
therefore, enters each $\eta$-vertex, one arrives at the following list of changes needed to proceed from the OPE potential to the OEE one:
(i) the flavour factor +1, for the OPE in the isovector channel, should be replaced by $-1$, for the OEE; (ii) the pion mass $m_\pi$ should be replaced by the $\eta$ mass 
$m_\eta$, and (iii) the pion coupling constant $g_b$ should be replaced by the $\eta$ coupling constant
$g_b/\sqrt3$.

Interestingly, if the flavour symmetry group $SU(3)$ is extended to $U(3)$ then the potentials from the $\pi$-, $\eta$-, and $\eta'$-exchange 
cancel each other identically in the strict flavour symmetry limit \cite{Aceti:2014kja}. 
However, because of the $U(1)$ anomaly, there are no reasons to expect the $\eta'$ to possess properties of the Goldstone boson 
\cite{DiVecchia:1980yfw, Rosenzweig:1979ay,Kawarabayashi:1980dp,Witten:1980sp}.

\end{document}